\documentclass[12pt]{article}
\pdfoutput=1
\usepackage{jheppub}
\usepackage{amsmath}
\usepackage{bm}
\usepackage{slashed}
\usepackage{graphicx}
\usepackage{makecell}

\newlength{\dummysp}
\newcommand{\beq}{\begin{eqnarray}}
\newcommand{\eeq}{\end{eqnarray}}

\newcommand{\gappeq}{\mathrel{\rlap {\raise.5ex\hbox{$>$}}
{\lower.5ex\hbox{$\sim$}}}}
\newcommand{\lappeq}{\mathrel{\rlap{\raise.5ex\hbox{$<$}}
{\lower.5ex\hbox{$\sim$}}}}

\newcommand{\ben}{\begin{enumerate}}
\newcommand{\een}{\end{enumerate}}

\newcommand{\bit}{\begin{itemize}}
\newcommand{\eit}{\end{itemize}}

\def\[{\left [}
\def\]{\right ]}
\def\({\left (}
\def\){\right )}

\def\Z{{\mathbb Z}}

\def\a{\alpha}
\def\b{\beta}

\def\k{\kappa}

\def\l{\lambda}

\def\m{\mu}

\def\Z{\mathbb{Z}}

\def\longra{\longrightarrow}

\def\bf{\textbf}
\def\it{\textit}
\def\f{\frac}

\title{$2$-index chiral gauge theories}
   
 \author{Mohamed M. Anber,}\author{Samson Y.L. Chan} 
  
\affiliation{Centre for Particle Theory, Department of Mathematical Sciences, Durham University, South Road, Durham DH1 3LE, UK}

\emailAdd{mohamed.anber@durham.ac.uk}\emailAdd{samson.y.chan@durham.ac.uk}    

\abstract{

{\flushleft{W}}e undertake a systematic study of the $4$-dimensional $SU(N)$ $2$-index chiral gauge theories and investigate their faithful global symmetries and dynamics.  These are a finite set of theories with fermions in the $2$-index symmetric and anti-symmetric representations, with no fundamentals, and they do not admit a large-$N$ limit. We employ a combination of perturbative and nonperturbative methods, enabling us to constrain their infrared (IR) phases. Specifically, we leverage the 't Hooft anomalies associated with continuous and discrete groups to eliminate a few scenarios. In some cases, the anomalies rule out the possibility of fermion composites. In other cases, the interplay between the continuous and discrete anomalies leads to multiple higher-order condensates, which inevitably form to match the anomalies. Further, we pinpoint the most probable symmetry-breaking patterns by searching for condensates that match the full set of anomalies resulting in the smallest number of IR degrees of freedom.   Higher-loop $\beta$-function analysis suggests that a few theories may flow to a conformal fixed point.}

\begin{document}

\maketitle

\flushbottom

\section{Introduction}

Chiral gauge theories form the fundamental framework of the Standard Model (SM) of particle physics. Within the SM, the electroweak sector undergoes Higgsing at weak coupling, allowing us to apply perturbative techniques. However, without a Higgs field, gauge theories generally flow towards a strongly-coupled regime, rendering their study considerably more challenging. A non-comprehensive list of some of the recent papers that studied chiral gauge theories is \cite{Bolognesi:2022beq,Bolognesi:2021yni,Bolognesi:2020mpe,Bolognesi:2017pek,Bai:2021tgl,Csaki:2021xhi,Csaki:2021aqv,Anber:2021iip,Smith:2021vbf,Sheu:2022odl}. 

This paper focuses on a class of $SU(N)$ chiral gauge theories that accommodate fermions in the $2$-index symmetric and anti-symmetric representations. These theories, referred to as $2$-index chiral gauge theories, can be characterized by the pair $(N,k)$, where $N$ represents the color and $k$ serves as a common divisor of $N+4$ and $N-4$. Moreover, $k$ is directly associated with the number of flavors in the $2$-index symmetric and anti-symmetric representations. What makes these theories particularly intriguing is the absence of a requirement to introduce fundamental fermions to cancel the gauge anomaly. Additionally, they exhibit non-asymptotic freedom for $N>44$. This class encompasses a collection of $14$ distinct theories, occupying a distinct region within asymptotically-free chiral gauge theories. Consequently, a systematic approach to studying this class is justified. It is divided into two subclasses: bosonic and fermion theories. The latter can accommodate gauge-invariant massless fermions. In comparison, the gauge-invariant operators in bosonic theories cannot have a spinor index, as the fermion number is gauged. 

We initiated this study in \cite{Anber:2021iip}, utilizing 't Hooft anomalies to constrain the infrared dynamics of two theories. Namely, these are $(N=8, k=4)$ and $(N=8,k=2)$ theories.  One important development was the identification of the faithful global symmetry acting on fermions. This enabled us to turn on the most general discrete fluxes, the color-flavor-$U(1)$ (CFU) fluxes compatible with the theory and, thus, utilize the full power of 't Hooft anomaly matching conditions. These anomalies are dubbed CFU anomalies. A theory with 't Hooft anomalies cannot be trivially gapped; the infrared (IR) spectrum must contain massless particles or multi-vacua. In the case $(N=8, k=4)$, we found that the condensation of two operators can saturate the anomalies. 

Continuing our explorations within this comprehensive framework, our current investigation exhausts all the $14$ theories and introduces a few novel aspects.
\begin{enumerate}

\item We incorporate anomalies stemming from discrete symmetries, thereby imposing additional constraints on the infrared spectra. In the context of the $2$-index chiral gauge theories we are examining, in addition to continuous non-abelian flavor symmetries, an axial $U(1)_A$ symmetry comes into play. When a bosonic operator condenses, it generally breaks the $U(1)_A$ symmetry down to a discrete subgroup, which typically is anomalous. Consequently, we face the challenge of identifying a set of condensates that not only matches the anomalies associated with non-abelian symmetries but also avoids the presence of any anomalous unbroken discrete subgroups. For example, the two candidate condensates we previously considered in the case $(N=8, k=4)$, \cite{Anber:2021iip}, fail to match the anomaly of an unbroken discrete group. We revise the situation in light of the new understanding and propose that the set of anomalies can be matched by other condensates. 

Interestingly, in a few cases, matching the full set of anomalies, particularly anomalies of discrete symmetries, can be achieved only via the condensation of multiple higher-order operators. Given the strong dynamics, such formation is not a surprise. However, anomalies explain the kinematical reasons why such condensates have to form. 

\item Another significant aspect of our work lies in our pursuit of the minimal scenario that satisfies the entire set of anomalies and yields the smallest number of massless particles in the infrared spectrum. Such a scenario holds particular appeal as it minimizes the free energy associated with the theory.  

\item We adhered to a systematic algorithm during our quest to identify composite massless fermions capable of satisfying the anomalies within the fermionic class. Regrettably, we were unable to find such composites. Notably, in the case of $(N=6, k=2)$, we demonstrate that these composites cannot solely match the CFU anomaly. Consequently, we are left with two plausible explanations: either these composites do not exist altogether, or the formation of condensates alongside the composites is necessary to match the anomaly. However, the latter scenario is rather contrived in that it requires the formation of special condensates (that do not alter the anomalies matched by the composites) in addition to the composite fermions, prompting us to lean toward the likelihood that composites cannot form in this case. 

\item To complete our study, we also examined the possibility that a theory flows to a conformal fixed point in the IR. Generally speaking, a theory can form a strongly-coupled IR fixed point, which is beyond the scope of any perturbative analysis. Such a fixed point automatically satisfies the full set of anomalies, albeit it remains an open question how this can be seen. Nevertheless, by scrutinizing the higher-order $\beta$-function, we successfully identified several instances where the analysis of the $\beta$-function offers indications suggestive of a perturbative nature inherent to a fixed point.

\end{enumerate}

This paper is organized as follows. In Section \ref{Theory: symmetry structure and anomalies}, we review the global symmetries and anomalies of the $2$-index chiral theories. This includes the anomalies of continuous symmetries, the CFU anomalies, as well as anomalies of discrete symmetries. In Section \ref{Anomaly matching and the IR phase}, we revise the matching of the CFU anomalies in the IR and introduce the novelty of matching the discrete subgroups of the axial $U(1)_A$ that can be left unbroken by a condensate. Sections \ref{Fermionic theories} and \ref{Bosonic theories} are devoted to applying these ideas to both the fermionic and bosonic field theories, respectively. We summarize our findings in Section \ref{Summary}, and in particular, the reader is referred to Table \ref{The IR realizations}, which, for all theories, it gives the global symmetries, the proposed IR condensates that yield the smallest number of Goldstones, and the fate of the symmetries in the IR.

\section{Theory: symmetry structure and anomalies}
\label{Theory: symmetry structure and anomalies}

We consider $SU(N)$ gauge theory with $n_\psi$ flavors of left-handed Weyl fermions $\psi$ transforming in the $2$-index representation along with $n_\chi$ flavors of left-handed Weyl fermions $\chi$ transforming in the complex conjugate $2$-index anti-symmetric representation:
\begin{eqnarray}
{\cal L}=-\frac{1}{2g^2}\mbox{tr}[f^{c}\wedge \star f^{c}] -i \bar \psi \bar \sigma^\mu D_\mu\psi -i\bar \chi \bar \sigma^\mu D_\mu \chi\,,
\end{eqnarray}
where $D_\mu\equiv \partial_\mu -i a_\mu^c$ is the covariant derivative, $a^c$ is the color gauge field, and $f^{c}$ is its field strength. In this work, we use the lower-case letters, $a^c$ and $f^c=da^c$, to denote the dynamical (color) $1$-form gauge field and its field strength, while we use upper-case letters, $A$ and $F=dA$, for background fields.  To keep track of the color indices, we choose $\psi_{(a_1a_2)}$ to carry two down indices, while $\chi_{[a_1a_2...a_{N-2}]}$ carries $N-2$ down indices. A round (square) bracket indicates symmetrizing (anti-symmetrizing) over the indices.  The cubic anomaly coefficients of the $2$-index symmetric and the conjugate of the $2$-index anti-symmetric representations are ${\cal A}_\psi=N+4$, ${\cal A}_{\chi}=-(N-4)$, respectively. Cancellation of the gauge anomaly demands that $n_\psi$ and $n_{\chi}$ are fixed as
\begin{eqnarray}
n_\psi=\frac{N-4}{k}\,, \quad n_{\chi}=\frac{N+4}{k}\,,
\end{eqnarray}
where $k$ is a common divisor of $N-4$ and $N+4$. The theory is asymptotically free provided that $11N-\frac{2(N^2-8)}{k}>0$. This leaves us with the finite set of theories in Table \ref{tab:2-index_chiral_theory}. 
These theories do not possess a large-$N$ limit, as they become infrared-free for $N>44$. Also, except for $N=5,6,10$, the other allowed colors are multiples of $4$. These are bosonic theories because all their gauge invariant operators cannot carry a spinor index. In other words, the $(-1)^F$ fermion number in bosonic theories is gauged, and thus,  they cannot have gauge-singlet fermionic operators. 

One important aspect of this work is to systematically analyze these theories, paying particular attention to the faithful global symmetries, and exhausting the class of generalized 't Hooft anomalies that enable us to constrain the infrared phases. 

\begin{table}
\center
\begin{tabular}{|c|c|c|c|c|c|c|c|c|c|c|}
\hline
$N$ &  $5$ &   $6$ &   $8$ &   $10$ &   $12$  &  $16$  &  $20$ &   $28$ &   $36$ &   $44$ \\
\hline
 $k$ & $1$ &  $1,2$ &  $2,4$ &  $2$ &  $4,8$ &  $4$ &  $4,8$ &  $8$ &  $8$ & $8$\\
\hline
\end{tabular}\,.
\caption{A list of the $2$-index chiral gauge theories.}
\label{tab:2-index_chiral_theory}
\end{table}
%

\subsection{Symmetries}

The theory enjoys two global flavor groups $SU(n_\psi)=SU((N-4)/k)$ and $SU(n_\chi)=SU((N+4)/k)$ acting on $\psi$ and $\chi$, respectively. 
In addition, the theory is endowed with two $U(1)$ global classical symmetries, $U(1)_1\times U(1)_2$. Their action on $\psi$ and $\chi$ is chosen as
\begin{eqnarray}
\nonumber
U(1)_1: \quad \psi\longrightarrow e^{i\alpha_1}\psi\,,\quad \chi\longrightarrow e^{i\beta_1}\chi\,,\\
U(1)_2: \quad \psi\longrightarrow e^{i\alpha_2}\psi\,,\quad \chi\longrightarrow e^{i\beta_2}\chi\,.
\end{eqnarray}
The two transformations $U(1)_1$ and $U(1)_2$ come naturally with two parameters. Here, however, we introduce the $4$ parameters $\alpha_{1,2}$ and $\beta_{1,2}$ to account for the fermions charges, in addition to the transformation parameters. 
 The gauge sector instantons break most of the classical $U(1)$ symmetries. The effective action in the instanton background acquires the terms
\begin{eqnarray}
\nonumber
\Delta S&=& i\left(n_\psi \alpha_1 T_\psi+n_\chi\beta_1 T_\chi\right)\int_{\mathbb M^4} \lambda_0^{u_1} \frac{\mbox{tr}\left[f^c\wedge f^c\right]}{8\pi^2}+i\left(n_\psi \alpha_2 T_\psi+n_\chi\beta_2 T_\chi\right)\int_{\mathbb M^4} \lambda_0^{u_2}  \frac{\mbox{tr}\left[f^c\wedge f^c\right]}{8\pi^2}\\
\end{eqnarray}
upon performing simultaneous transformations of $U(1)_1\times U(1)_2$, where
\begin{eqnarray}
T_{\psi}=N+2\,,\quad T_\chi=N-2
\end{eqnarray}
 are the Dynkin indices of the representations. Here, $f^c$ is the $2$-form field strength of the color group, while $\lambda^{u_1}$ and $\lambda^{u_2}$ are the gauge parameters of $U(1)_1$ and $U(2)_2$, respectively, i.e., $A^{u_{1,2}}\longrightarrow A^{u_{1,2}}+d\lambda^{u_{1,2}}$.  We can find a combination of the parameters $\alpha_1$ and $\beta_1$ that kills the first term in $\Delta S$, leaving behind a genuine symmetry. We call this symmetry the axial $U(1)_A$. It acts on $\psi$ and $\chi$ with transformation parameter $\a$:  
\begin{eqnarray}
U(1)_A:\quad \psi \longrightarrow e^{i2\pi\alpha q_\psi}\psi\,, \quad \chi\longrightarrow e^{i2\pi\alpha q_\chi}\chi\,,
\end{eqnarray}
and we have defined the $U(1)_A$ charges of $\psi$ and $\chi$ as
\begin{eqnarray}
q_\psi\equiv-\frac{ N_\chi}{r}\,, \quad q_\chi \equiv \frac{N_\psi}{r}\,,
\end{eqnarray}
where $r=\mbox{gcd}(n_\chi T_\chi, n_\psi T_\psi)$, and
\begin{eqnarray}
N_\chi\equiv n_\chi T_\chi\,,\quad N_\psi\equiv n_\psi T_\psi\,.
\end{eqnarray}

Yet, we can find values of $\alpha_2$ and $\beta_2$ that leave the discrete subgroup $\mathbb Z_{p_\psi N_\psi+ p_\chi N_\chi}\subset U(1)_2$ invariant in the color background, where $p_\psi$ and $p_\chi$ are arbitrary integers. In Appendix \ref{Obtaining the Discrete Chiral Symmetry}, we show that most of the $\mathbb Z_{p_\psi N_\psi+ p_\chi N_\chi}$ elements belong to $U(1)_A$ and that only a subgroup $\mathbb Z_r\subset\mathbb Z_{p_\psi N_\psi+ p_\chi N_\chi}$, which is $p_\chi$ and $p_\psi$-independent, can {\em potentially} act as a {\em genuine} symmetry on the fermions. Also, we can always choose $\mathbb Z_r$ to act solely on $\chi$:
\begin{eqnarray}
\mathbb Z_r:\quad (\psi,\chi)\longrightarrow \left(\psi, e^{i\frac{2\pi \ell}{r}} \chi\right)\,, \quad \ell=0,1,2,..., r-1\,.
\end{eqnarray}
Yet, one must check that $\mathbb Z_r$ or a subgroup of it cannot be absorbed in the centers of the color or flavor groups, which leaves a proper subgroup of $\mathbb Z_r$ as the genuine discrete symmetry.  This will be checked on a case-by-case basis. In the following, we will use $\mathbb Z_p^{d\chi}\subseteq \mathbb Z_r$ to denote the genuine discrete chiral symmetry. For completeness, we remind the reader that the fermion number symmetry $\mathbb Z_2^F=(-1)^F$ operates on $\psi$ and $\chi$ as $(\psi,\chi)\longrightarrow -(\psi, \chi)$.

  Finally, we also note that when $N$ is even, the theory is endowed with a $\mathbb Z^{(1)}_2$ $1$-form center symmetry acting on the fundamental Wilson loops.  In summary, the good global symmetry of the theory is 
\begin{eqnarray}
G^{\mbox{g}}\sim SU(n_\psi)\times SU(n_\chi)\times U(1)_A\times \mathbb Z_p^{d\chi}\times \mathbb Z_{\scriptsize\mbox{gcd}(N,2)}^{(1)} \,,
\label{global group mod discrete}
\end{eqnarray}
where the tilde indicates that this is the correct group modulo a discrete group needed to fix the faithful global symmetry. Thus, the faithful global symmetry is a quotient group.

To determine the correct quotient group, we follow  \cite{Anber:2019nze, Anber:2021iip}. Here, we keep our treatment short as the details can be found in \cite{Anber:2021iip}. We put the theory on a general compact $4$-D spin manifold $\mathbb M^4$, define a principal bundle of the continuous part of the global symmetry $G^{\scriptsize\mbox{g}}$  on $\mathbb M^4$, and take the transition functions of $G^{\scriptsize\mbox{g}}$ to act on fibers by left multiplication. Spinors are sections of the bundle, and we use the notations $\psi_i$ and $\chi_i$ for their values on a local patch $U_i\subset \mathbb M^4$.  We denote the transition functions of the color $SU(N)$, (non-abelian) flavor, and $U(1)_A$ group as $g$, $f$, and $u$, respectively, along with the proper superscript to distinguish those of $\psi$ and $\chi$. 
On the overlap $U_i \cap U_j$ we have
\beq
\psi_i = (g^\psi, f^\psi, u^\psi)_{ij} \, \psi_j~,\quad \chi_i = (g^\chi, f^\chi, u^\chi)_{ij} \, \chi_j~.
\eeq
The fermions are well defined on $\mathbb M^4$ provided they satisfy the cocycle condition (a necessary consistency condition) on the triple overlap $U_i \cap U_j\cap U_k$.
Now,  we turn on background gauge fields for centers of the gauge, flavor, and $U(1)_A$ groups and determine the most general combination compatible with the cocycle condition \footnote{See \cite{Anber:2020qzb,Anber:2023urd,Anber:2020xfk,Nakajima:2022jxg} for applications of the anomalies resulting from turning on these fluxes.}, which reads
\beq
\left( g^\psi, f^\psi, u^\psi \right)_{ij} \circ \left( g^\psi, f^\psi, u^\psi \right)_{jk} \circ \left( g^\psi, f^\psi, u^\psi \right)_{ki} = \left( z_c, z_f, z_u \right) \quad {\rm with} \quad  z_c z_f z_u = 1~,
\eeq
where $z$'s refer to the center elements: $z_c \in \mathbb{Z}_{N/{\rm gcd}(N,2)}$, $z_f \in \mathbb{Z}_{n_\psi}$, and $z_u \in U(1)_A$.
The condition $z_c z_f z_u = 1$ is required for the equivalence relation
\beq
\left( z_c, z_f, z_u \right) \sim (1,1,1)~,
\label{eq:equivalence relation}
\eeq
which is needed to obtain the correct compatibility condition. Similar expressions hold for the cocycle condition of $\chi$. The following two equations give the consistency (compatibility) conditions
\begin{eqnarray}
\nonumber
\psi&:&\quad \underbrace{e^{i 2\pi \frac{2m}{N}}}_{z_c}\underbrace{e^{i 2\pi\frac{pk}{N-4}}}_{z_\psi}\underbrace{e^{-i2\pi s\frac{(N+4)(N-2)}{kr}}}_{z_u}=1\,,\\
\chi&:&\quad \underbrace{e^{-i 2\pi \frac{2m}{N}}}_{z_c}\underbrace{e^{-i 2\pi\frac{p'k}{N+4}}}_{z_\chi}\underbrace{e^{i2\pi s\frac{(N-4)(N+2)}{kr}}}_{z_u}=1\,.
\label{consistency conditions}
\end{eqnarray} 
 Here, $m \in \mathbb Z_{\scriptsize N/\mbox{gcd}(N,2)}$, $p\in \mathbb Z_{n_\psi}$, $p'\in \mathbb Z_{n_{\chi}}$ and $s$ is a $U(1)_A$ parameter. The factor of $2$ that appears in $z_c$ accounts for the $N$-ality of $\psi$ and $\chi$, and the negative sign that appears in the $z_c$ factor in the second line accounts for the fact that $\chi$ transforms in the complex conjugate of the $2$-index anti-symmetric representation. We also take $\psi$ to transform in the fundamental representation of $SU(n_\psi)$ and $\chi$ to transform in the anti-fundamental representation of $SU(n_\chi)$. Following \cite{Anber:2021iip}, we shall dub the discrete color-flavor-$U(1)_A$ fluxes as the CFU fluxes. The full set of solutions of (\ref{consistency conditions}) determines the quotient group in (\ref{global group mod discrete}). These solutions will be found on a case-by-case basis. In general, we divide (\ref{global group mod discrete}) by $\mathbb Z_{\scriptsize N/\mbox{gcd}(N,2)}\times \mathbb Z_{(N-4)/k}\times \mathbb Z_{(N+4)/k}$ or a subgroup of it.

Once a non-trivial solution of (\ref{consistency conditions}) is found, we can calculate the topological charges associated with the center fluxes, which are fractional charges in general. Let $\mathbb M^4$ admit two independent $2$-cycles and let two integers, e.g., $m_1$ and $m_2$, account for the number of quanta piercing through them. For example, we can take $\mathbb M^4=\mathbb T^4$, a $4$-torus with a period length $L$, and turn on the gauge fields that are compatible with the cocycle condition:
\begin{eqnarray}
\nonumber
a_1^c&=&\frac{2\pi m_1}{L^2}\bm H_c\cdot \bm \nu_c\,,\quad a_2^c=0\,, \quad a_3^c=\frac{2\pi m_2}{L^2}\bm H_c\cdot \bm \nu_c\,,\quad a_4^c=0\,,\\
\nonumber
A_1^\psi&=&\frac{2\pi p_1}{L^2}\bm H_\psi\cdot \bm \nu_\psi\,,\quad A_2^\psi=0\,, \quad A_3^\psi=\frac{2\pi p_2}{L^2}\bm H_\psi\cdot \bm \nu_\psi\,,\quad A_4^\psi=0\,,\\
\nonumber
A_1^\chi&=&\frac{2\pi p_1'}{L^2}\bm H_\chi\cdot \bm \nu_\chi\,,\quad A_2^\chi=0\,, \quad A_3^\chi=\frac{2\pi p_2'}{L^2}\bm H_\chi\cdot \bm \nu_\chi\,,\quad A_4^\chi=0\,,\\
A_1^u&=&\frac{2\pi s_1}{L^2}\,,\quad A_2^u=0\,, \quad A_3^u=\frac{2\pi s_2}{L^2}\,,\quad A_4^u=0\,.
\label{background fields}
\end{eqnarray}
$a_\mu^c$, $A_\mu^\psi$, $A_\mu^\chi$, and $A^u_\mu$ are the background gauge fields of the center of the color, $SU(n_\psi)$ flavor, $SU(n_\chi)$, flavor, and $U(1)_A$, respectively. The bold-face letters $\bm H\equiv (H_1,...,H_{N-1})$ are the Cartan generators of $SU(N)$ group, while $\bm\nu\equiv(\nu_1,...,\nu_{N-1})$ is a weight in the defining representation of the group.  Notice that the integers $m_{1,2},p_{1,2},p'_{1,2},s_{1,2}$ are the same integers that solve the consistency conditions (\ref{consistency conditions}). Given the set of the background fields (\ref{background fields}), one immediately obtains the topological charges defined as
\begin{eqnarray}
\nonumber
Q_c=\int_{\mathbb T^4}\frac{\mbox{tr}\left[f^c\wedge f^c\right]}{8\pi^2}\,,Q_\psi=\int_{\mathbb T^4}\frac{\mbox{tr}\left[F^\psi\wedge F^\psi\right]}{8\pi^2}\,, Q_\chi=\int_{\mathbb T^4}\frac{\mbox{tr}\left[F^\chi\wedge F^\chi\right]}{8\pi^2}\,, Q_u=\int_{\mathbb T^4}\frac{F^u\wedge F^u}{8\pi^2}\,,\\
\label{the topological charges in general}
\end{eqnarray}
and $f^c$, $F^{\psi,\chi,u}$ are the field strengths of the corresponding background.
Substituting (\ref{background fields}) into (\ref{the topological charges in general}), we obtain
\begin{eqnarray}
\nonumber
Q_c&=&k_c-\frac{m_1m_2}{N}\,,\quad Q_\psi=k_\psi-\frac{ p_1p_2k}{N-4}\,,\\
 \quad Q_\chi&=&k_\chi-\frac{  p'_1p'_2k}{N+4}\,,\quad Q_u=(s_1-k_1)(s_2-k_2)\,,
 \label{fractional topological charges}
\end{eqnarray}
and $k_c, k_\psi,k_\chi,k_1,k_2 \in \mathbb Z$ are arbitrary integers that are always allowed.
These fluxes will support fermion zero modes, and the Dirac indices give their number:
\begin{eqnarray}
\nonumber
{\cal I}_\psi &=&n_\psi T_\psi Q_c+\mbox{dim}_\psi Q_\psi+\mbox{dim}_\psi n_\psi q_\psi^2 Q_u\,,\\
{\cal I}_\chi &=&n_\chi T_\chi Q_c+\mbox{dim}_\chi Q_\chi+\mbox{dim}_\chi n_\chi q_\chi^2 Q_u\,,
\label{Dirac indices}
\end{eqnarray}
and $\mbox{dim}_\psi=\frac{N(N+1)}{2}$, $\mbox{dim}_\chi=\frac{N(N-1)}{2}$ are the dimensions of the representations. Dirac indices count the number of the Weyl zero modes in the background of center fluxes. The integrality of the indices can work as a check on the consistency of the fluxes on $\mathbb M^4$. 

One may also turn on the CFU fluxes on nonspin $\mathbb M^4$. A nonspin manifold does not admit fermions in the sense that there is an obstruction in lifting the $SO(4)$ rotation group bundle to a $\mbox{Spin}(4)$ bundle on $\mathbb M^4$. A diagnosis of a non-spin manifold is that the Dirac index of a Weyl fermion, ${\cal I}=\frac{1}{196\pi^2}\int_{\mathbb M^4}\mbox{tr}R\wedge R$, where $R$ is the curvature $2$-form, is non-integer. An example of a nonspin manifold is $\mathbb {CP}^2$, which has $\frac{1}{196\pi^2}\int_{\mathbb {CP}^2}\mbox{tr}R\wedge R=-\frac{1}{8}$ . To put a Weyl fermion on $\mathbb {CP}^2$, we need to excite a $U(1)$ flux $F$ through its $2$-cycle $\mathbb {CP}^1\subset \mathbb {CP}^2$ and demand that $\int_{\mathbb {CP}^1}F\in \pi (2\mathbb Z+1)$, which implies $\frac{1}{8\pi^2}\int_{\mathbb {CP}^2}F\wedge F\in\frac{\mathbb Z}{8}$. Now, one can easily check the integrality of the Dirac index $\frac{1}{196\pi^2}\int_{\mathbb {CP}^2}\mbox{tr}R\wedge R+\frac{1}{8\pi^2}\int_{\mathbb {CP}^2}F\wedge F \in \mathbb Z$, and thus, the fermions are well-defined on $\mathbb {CP}^2$ in the presence of such $U(1)$ fluxes. Here, although one cannot define a $\mbox{Spin}(4)$ bundle on pure $\mathbb {CP}^2$, in the sense that the corresponding cocycle condition fails on a triple overlap, nonetheless, we can define the $\mbox{Spin}_c(4)$ structure $\mbox{Spin}(4)\times U(1)/\mathbb Z_2$ in the presence of the $U(1)$ background.  

This idea can be generalized in the presence of the CFU fluxes; see \cite{Anber:2020gig} for details. One just needs to replace the consistency conditions (\ref{consistency conditions})  with
\begin{eqnarray}
\nonumber
\psi&:&\quad \underbrace{e^{i 2\pi \frac{2m}{N}}}_{z_c}\underbrace{e^{i 2\pi\frac{pk}{N-4}}}_{z_\psi}\underbrace{e^{-i2\pi s\frac{(N+4)(N-2)}{kr}}}_{z_u}=-1\,,\\
\chi&:&\quad \underbrace{e^{-i 2\pi \frac{2m}{N}}}_{z_c}\underbrace{e^{-i 2\pi\frac{p'k}{N+4}}}_{z_\chi}\underbrace{e^{i2\pi s\frac{(N-4)(N+2)}{kr}}}_{z_u}=-1\,.
\label{consistency conditions nonspin}
\end{eqnarray} 
The minus sign on the right-hand side compensates for the minus sign arising from parallel transporting the spinor fields around appropriate closed paths in $\mathbb {CP}^2$; see the detailed discussion in \cite{Anber:2020gig}. Given that a solution, $m \in \mathbb Z_{\scriptsize N/\mbox{gcd}(N,2)}$, $p\in \mathbb Z_{n_\psi}$, $p'\in \mathbb Z_{n_{\chi}}$ and $s$, to (\ref{consistency conditions nonspin}) can be found, the topological charges corresponding to the CFU fluxes and gravity are given by (see \cite{Anber:2020gig})
\begin{eqnarray}
\nonumber
Q_{c}&=\frac{m^2}{2}\left(1-\frac{1}{N}\right)~, \quad Q_{\psi}=\frac{p^2}{2}\left(1-\frac{k}{N-4}\right)~,
\\[3pt]
Q_{\chi}&=\frac{p'^2}{2}\left(1-\frac{k}{N+4}\right)~, \quad Q_{u}=\frac{1}{2}s^2\,, \quad Q_g=-\frac{1}{8}~\,.
\end{eqnarray}
The Dirac-indices of $\psi$ and $\chi$ are 
\begin{eqnarray}
\nonumber
{\cal I}_{\psi}^{\mathbb {CP}^2}&=&n_\psi T_{\psi}Q_{c}+ \mbox{dim}_\psi  Q_{\psi}+\mbox{dim}_\psi n_\psi  \left(q_\psi^2 Q_{u}+Q_g\right)~,
\\[3pt]
{\cal I}_{\chi}^{\mathbb {CP}^2}&=&n_\chi T_{\chi}Q_{c}+ \mbox{dim}_{\chi}  Q_{\chi}+\mbox{dim}_{\chi} n_\chi  \left(q_\chi^2 Q_{u}+Q_g\right)~,
\label{nonspin Dirac indices}
\end{eqnarray}
which are always integers. Except for $(N=6,k=2)$ and $(N=10,k=2)$  in Table \ref{tab:2-index_chiral_theory}, we can always find solutions to (\ref{consistency conditions nonspin}), and thus, we can put these theories on $\mathbb {CP}^2$.

\subsection{Anomalies}

The theory has a set of 't Hooft anomalies that can help constrain the possible IR phases.  In the following, we list the 't Hooft anomalies we shall encounter in our study. 

\subsection*{ (I) $\left[SU(n_\psi)\right]^3$ and $\left[SU(n_\chi)\right]^3$ anomalies }

These are perturbative (triangle) anomalies and their inflow from $5$-D to $4$-D is captured via $5$-D Chern-Simons theories:
\begin{eqnarray}
\nonumber
\left[SU(n_\psi)\right]^3&:& \quad \exp\left[i\, \mbox{dim}_\psi \int_{{\mathbb M}^5}\omega_5(A^\psi) \right]\,,\\
\left[SU(n_\chi)\right]^3&:& \quad \exp\left[i\, \mbox{dim}_\chi \int_{{\mathbb M}^5}\omega_5(A^\chi) \right]\,,
\end{eqnarray}
where $A^\psi$ and $A^\chi$ are the $SU(n_\psi)$ and $SU(n_\chi)$ $1$-form background gauge fields, extended from $4$-D to $5$-D, and $\omega_5(A)$ is the $5$-D Chern-Simons form defined via the descent equation:
\begin{eqnarray}
d\omega_5(A)=\frac{1}{3! (2\pi)^2}\mbox{tr}_\Box F^3\,,
\end{eqnarray}
and $F$ is the $2$-form field strength of $A$. 

\subsection*{(II) $U(1)_A$- and $\mathbb Z_p^{d\chi}$-gravitational anomalies }

These anomalies are captured via the $5$-D anomaly inflow actions:
\begin{eqnarray}
\nonumber
U(1)_A[\mbox{grav}]&:& \quad \exp\left[ i\left(q_\psi n_\psi \mbox{dim}_\psi+q_\chi n_\chi \mbox{dim}_\chi\right)\int_{\mathbb M^5} A^u\wedge
\frac{p_1 (\mathbb M^5)}{24} \right]\,,\\
\mathbb Z_p^{d\chi}[\mbox{grav}]&:& \quad \exp\left[ i\left(n_\chi \mbox{dim}_\chi\right)\int_{\mathbb M^5} A^{d\chi}\wedge
\frac{p_1 (\mathbb M^5)}{24} \right]\,.
\label{gravity anomaly}
\end{eqnarray}
The $1$-form gauge fields $A^u$ and $A^{d\chi}$ are the backgrounds of $U(1)_A$ and $\mathbb Z_p^{d\chi}$, respectively.  $p_1 (\mathbb M^5)=-\frac{1}{8\pi^2}R\wedge R$ is the first Pontryagin number and $R$ is the curvature $2$-form. On a spin manifold, we have $\int_{\mathbb M^4}p_1(\mathbb M^4)\in 48\,\mathbb Z$, and thus, there are $2$ zero modes per Weyl fermion in a gravitational background.  Under $U(1)_A$ and $\mathbb Z_p^{d\chi}$ transformations we have $A^u\longrightarrow A^u+d\lambda^u$ with $\oint d\lambda^u= 2\pi \mathbb Z$ and $A^{d\chi}\longrightarrow A^{d\chi}+d\lambda^{d\chi}$, with $\oint d\lambda^{d\chi}=\frac{2\pi \mathbb Z}{p}$,  and the anomaly inflow actions produce the $4$-D anomalies.

The result (\ref{gravity anomaly}) is ``perturbative'' as it can be seen from a triangle diagram with two vertices that couple the fermions to a gravitational background via the energy-momentum tensor, while the third vertex couples the fermions to an external $U(1)_A$ or $\mathbb Z_{p}^{d\chi}$ sources. 

\subsection*{(III) CFU anomalies}

These anomalies were identified in \cite{Anber:2019nze}; however, see \cite{Tanizaki:2018wtg,Shimizu:2017asf} for earlier encounters. They are anomalies of $U(1)_A$ and $\mathbb Z_p^{d\chi}$ symmetries in the background of the CFU fluxes that are supported on a general spin manifold. As was shown in \cite{Anber:2021iip}, $5$-D anomaly inflow actions can also capture them. However, we find it more convenient to express such anomalies in terms of the non-trivial phases that are acquired by the partition function ${\cal Z}$ under the action of  $U(1)_A$ and $\mathbb Z_p^{d\chi}$ symmetries in the background of the CFU fluxes:
\begin{eqnarray}
\nonumber
U(1)_A[\mbox{CFU}]&:&  {\cal Z}\longrightarrow e^{i 2\pi \alpha \left( q_\psi {\cal I}_\psi+ q_\chi {\cal I}_\chi\right)} {\cal Z}\,,\\
\mathbb Z_p^{d\chi}[\mbox{CFU}]&:&  {\cal Z}\longrightarrow e^{i \frac{2\pi}{p}  {\cal I}_\chi} {\cal Z}\,,
\label{CFU anomaly spin}
\end{eqnarray}
and the Dirac indices $ {\cal I}_\psi$ and $ {\cal I}_\chi$ are given in (\ref{Dirac indices}).  The contribution from the color topological charge $Q_c$ drops out in the computation of the $U(1)_A[\mbox{CFU}]$ anomaly, as can be easily checked, since $U(1)_A$ is a good symmetry in the background of the color flux. This is not the case with $\mathbb Z_p^{d\chi}[\mbox{CFU}]$ anomaly, where $Q_c$ contributes to the anomaly. As we shall discuss,  this observation has important consequences for anomaly matching in the IR.
 
  It is also important to notice that the perturbative anomalies $U(1)_A[SU(n_\psi)]^2$,  $U(1)_A[SU(n_\chi)]^2$, $\mathbb Z_p^{d\chi}[SU(n_\chi)]^2$, and $[U(1)_A]^3$ are a subset of the CFU anomalies, obtained by turning off the center fluxes and keeping only the integer topological charges in (\ref{fractional topological charges}). Again, one can express them using anomaly inflow actions as
\begin{eqnarray}
\nonumber
U(1)_A[SU(n_\psi)]^2&:&\quad  \exp\left[ i q_\psi n_\psi \mbox{dim}_\psi \int_{\mathbb M^5} A^u\wedge
\frac{\mbox{tr}\left[F^\psi\wedge F^\psi\right]}{8\pi^2} \right]\,,\\
\nonumber
U(1)_A[SU(n_\chi)]^2&:&\quad  \exp\left[ i q_\chi n_\chi \mbox{dim}_\chi \int_{\mathbb M^5} A^u\wedge
\frac{\mbox{tr}\left[F^\chi\wedge F^\chi\right]}{8\pi^2} \right]\,,\\
\nonumber
\mathbb Z_p^{d\chi}[SU(n_\chi)]^2&:&\quad  \exp\left[ i n_\chi \mbox{dim}_\chi \int_{\mathbb M^5} A^{d\chi}\wedge
\frac{\mbox{tr}\left[F^\chi\wedge F^\chi\right]}{8\pi^2} \right]\,,\\
 \left[U(1)_A\right]^3&:&\quad \exp\left[ i \left(q_\psi^3n_\psi \mbox{dim}_\psi+q_\chi^3n_\chi \mbox{dim}_\chi\right) \int_{\mathbb M^5} A^{u}\wedge
\frac{F^u\wedge F^u}{24\pi^2} \right]\,.
\end{eqnarray}
In addition, for $N$ even, the CFU anomalies encompass the $U(1)_A$ $0$-form/ $\mathbb Z_2^{[1]}$ $1$-form as well as the $\mathbb Z_r^{d\chi}$ $0$-form/ $\mathbb Z_2^{[1]}$ $1$-form mixed anomalies. These can be easily found by turning off the flavor and the $U(1)_A$ fluxes. In practice, one uses (\ref{CFU anomaly spin}), (\ref{fractional topological charges}),  and (\ref{Dirac indices}), after setting $p_{1,2}=p'_{1,2}=s_{1,2}=0$ and $m_1=m_2=N/2$. This choice enforces the consistency conditions (\ref{consistency conditions}) and gives $Q_\psi=Q_\chi=Q_u=0$ and $Q_c=\frac {N}{4}$.

One may also use the CFU fluxes on $\mathbb {CP}^2$ to calculate the $U(1)_A \left[\mbox{CFU}\right]$ and $\mathbb Z_{p} \left[\mbox{CFU}\right]$ anomalies, which sometimes are more restrictive than the corresponding anomalies on a spin manifold. We use the Dirac indices on ${\mathbb {CP}^2}$, as given by (\ref{nonspin Dirac indices}), to find
\beq
\nonumber
U(1)_A \left[\mbox{CFU}\right]_{\mathbb {CP}^2}&:& {\cal Z}\longrightarrow e^{i 2\pi \alpha\left[ q_\psi {\cal I}_{\psi}^{\mathbb {CP}^2} + q_\chi {\cal I}_{\chi}^{\mathbb {CP}^2}\right]}{\cal Z}~, \label{eq:CFU anomalies U(1)_R CP2}
\\[3pt]
\mathbb Z_{p}^{d\chi} \left[\mbox{CFU}\right]_{\mathbb {CP}^2}&:& {\cal Z}\longrightarrow e^{i \frac{2\pi}{p}{\cal I}_{\chi}^{\mathbb {CP}^2}}{\cal Z}~.
\label{eq:CFU anomalies Z_p CP2}
\eeq

From here on, we shall write all anomalies in terms of their phases to reduce clutter. For example, instead of (\ref{CFU anomaly spin}), we write:
\begin{eqnarray}
U(1)_A[\mbox{CFU}]:  q_\psi {\cal I}_\psi+ q_\chi {\cal I}_\chi\,,\quad
\mathbb Z_p^{d\chi}[\mbox{CFU}]:  {\cal I}_\chi\,.
\label{ shortCFU anomaly spin}
\end{eqnarray}
%

 \subsection*{(IV) Anomalies of discrete groups}

 Here, we consider anomalies of a discrete symmetry $\mathbb Z_n$, where $n$ is a general positive integer. An example of the discrete symmetry is the $\mathbb Z_{p}^{d\chi}$ chiral symmetry or a discrete subgroup of $U(1)_A$ left unbroken in the IR. The proper way to detect anomalies of discrete symmetries is to use the Dai-Freed prescription \cite{Dai:1994kq,Witten:2019bou}. The idea stems from the fact that a chiral massless fermion defined on $\mathbb M^4$ can be realized as the chiral zero mode residing on the boundary $\mathbb M^4$ of a  $5$-dimensional manifold $\mathbb M^5$ that is endowed with massive fermions, with $\mathbb Z_n$ turned on in the $5$-dimensional manifold. One can also consider a different $5$-dimensional manifold $\mathbb M'^5$ with the same boundary $\mathbb M^4$. If the partition functions defined on $\mathbb M^5$ and $\mathbb M'^5$ have the same phase, then the theory on $\mathbb M^4$ is uniquely defined and anomaly-free; otherwise, it is anomalous. Applying the Dai-Freed prescription to study the IR phases of strongly-coupled theories is innovative. However, see \cite{Razamat:2020kyf,Anber:2023urd} for previous applications\footnote{Also, see \cite{Garcia-Etxebarria:2018ajm,Davighi:2019rcd} for applications of Dai-Freed anomalies in particle physics.}.

If $\mathbb M^4$ is a spin manifold, the geometrical obstruction of uniquely extending a $4$-D theory to a $5$-D bulk can be inferred by computing the bordism group $\Omega_5^{\scriptsize \mbox{Spin}}(B\mathbb Z_n)$, where $B\mathbb Z_n$ is the classifying space of $\mathbb Z_n$\footnote{A classifying space of symmetry $G$ is an infinite dimensional space with the property that any principal $G$-bundle on a manifold $\mathbb M$ is the pullback via some map $f:\mathbb M\longrightarrow BG$. Then, the set of topologically distinct principal $G$-bundles over $\mathbb M$ is equivalent to the set of the homotopy classes of maps from $\mathbb M$ to $BG$.}.  If $\Omega_5^{\scriptsize \mbox{Spin}}(B\mathbb Z_n)$ is non-trivial, the theory might have a nonperturbative anomaly.  To find the anomaly, one computes the $\eta$-invariant, a resolvent of the spectral asymmetry of the Dirac operator, on specific closed $5$-dimensional spin manifolds that can detect the anomaly.   For example, one puts the theory on Lens spaces to gauge $\mathbb Z_n$ and discover whether the theory exhibits a nonperturbative anomaly. For $n$ even, $n=2m$, one can take $\mathbb M^4$ to be nonspin by employing the twisted symmetry group $\mbox{Spin}^{\mathbb Z_{2m}}(4)=(\mbox{Spin}(4)\times \mathbb Z_{2m})/\mathbb Z_2$ instead of $\mbox{Spin}(4)$. Here, one needs to compute the $\eta$-invariant that detects the bordism group $\Omega_5^{\scriptsize \mbox{Spin}^{\mathbb Z_{2m}}}$.

The computations of the relevant $\eta$-invariants were carried out in \cite{Hsieh:2018ifc} (see also \cite{Davighi:2020uab} for an alternative perspective).  For a theory of a left-handed Weyl fermion with a charge $s$ under $\mathbb Z_n$ defined on a spin manifold, the anomaly is given by the pair of phases: 
\begin{eqnarray}
\left\{ (n^2+3n+2)s^3 \, \mbox{mod}\, 6n\,, 2s\, \mbox{mod}\, n\right\}\,.
\label{nonperturbative discrete anomaly}
\end{eqnarray}
This pair can be thought of as contributions from $[\mathbb Z_n]^3$ and mixed $\mathbb Z_n\left[\mbox{grav}\right]$ anomalies. Indeed, the second entry in  (\ref{nonperturbative discrete anomaly}) is precisely the anomaly we computed in the second line of (\ref{gravity anomaly}). The first entry can be obtained from pure $[U(1)]^3$ anomaly by restricting $U(1)$ to a $\mathbb Z_n$ discrete group; this is the  Ibanez-Ross anomaly we comment on below. 
Also, a Weyl fermion defined on a twisted background and carrying a charge $s$ under $\mathbb Z_{2m}$ has an anomaly given by the pair\footnote{More generally, the bordism group $\Omega_5^{\scriptsize \mbox{Spin}^{\mathbb Z_{2m}}}\cong\mathbb Z_a\times \mathbb Z_b$, where $a,b$, and the associated anomalies are given by Eqs. (2.11)- (2.13) in \cite{Anber:2023urd}. In the present work, Eq. (\ref{twisted nonperturbative discrete anomaly}) suffices to tackle the theories at hand.}
\begin{eqnarray}
\left\{ \left((2m^2+m+1)s^3-(m+3)s\right) \, \mbox{mod}\, 48m\,, \left(ms^3+s\right)\, \mbox{mod}\, 2m\right\}\,.
\label{twisted nonperturbative discrete anomaly}
\end{eqnarray}
The charge $s$ is assumed to be odd such that the fermion transforms under the $\mathbb Z_2$ subgroup of $\mathbb Z_{2m}$.
Generally, the anomaly (\ref{twisted nonperturbative discrete anomaly}) is more restrictive than (\ref{nonperturbative discrete anomaly}). We shall use both (\ref{twisted nonperturbative discrete anomaly}) and (\ref{nonperturbative discrete anomaly}) to constrain our theories.  It is important to note that $\mathbb Z_2$ symmetry is anomaly-free, as can be easily seen from (\ref{nonperturbative discrete anomaly}) and (\ref{twisted nonperturbative discrete anomaly}). This observation will play an essential role in the IR anomaly matching by condensates, as many of them break the global symmetries to $\mathbb Z_2$, as we discuss below. 

We also comment on the Ibanez-Ross anomaly-matching conditions \cite{Ibanez:1991hv}. These are obtained from $[U(1)]^3$ and $U(1)[\mbox{grav}]$ anomalies by restricting $U(1)$ to a $\mathbb Z_n$ subgroup. The Ibanez-Ross anomaly-cancellation conditions read:
\begin{eqnarray}
s^3=p'n+\frac{r' n^3}{8}\,, \quad s=p''n+r''n\,,
\label{Ibanez-Ross}
\end{eqnarray}
where $p',r',p'',r''\in \mathbb Z$, $p'\in 3\mathbb Z$ if $n\in 3\mathbb Z$, and $r',r''=0$ if $n$ is odd. It can be shown that (\ref{Ibanez-Ross}) and (\ref{nonperturbative discrete anomaly}) are equivalent \cite{Hsieh:2018ifc}. Hence, in what follows, we use either (\ref{nonperturbative discrete anomaly}) or (\ref{twisted nonperturbative discrete anomaly}) to calculate discrete anomalies. 

Finally, we also may have a discrete anomaly of the form $\mathbb Z_m [\mathbb Z_n]^2$. Such an anomaly can descend from $U(1)_A[\mbox{CFU}]$ anomaly after a given condensate breaks $U(1)_A$ down to a discrete subgroup. Let $s$ and $s'$ be the charges of a left-handed Weyl fermion under $\mathbb Z_n$ and $\mathbb Z_m$, respectively. Then, the anomaly cancelation condition, which follows from the Ibanez-Ross conditions, is given by \cite{Csaki:1997aw}
\begin{eqnarray}
s^2s'=p'\, \mbox{gcd}(m,n)+\frac{p''}{8}mn^2\,,
\label{Zm Zn square anomaly}
\end{eqnarray}
where $p',p''\in \mathbb Z$ and $p''$ can be non-vanishing only if $n$ and $m$ are even. Notice that this anomaly is trivial when $\mathbb Z_m=\mathbb Z_2$.

\section{Anomaly matching and the IR phase}
\label{Anomaly matching and the IR phase}

\subsection{IR anomaly matching}

't Hooft anomalies preclude a trivially gapped IR phase: a theory with 't Hooft anomaly must have gapless excitations, degenerate vacua, or symmetry-preserving topological quantum field theory (TQFT).  In this work, we use the safe assumption that the gauge group does not break spontaneously under its strong dynamics. The breaking of a gauge group is under control in the presence of scalars at weak coupling, an ingredient absent from our theory from the get-go\footnote{Tumbling, \cite{Raby:1979my}, is a mechanism by which the breakdown of a gauge group
occurs without the aid of fundamental scalar fields. We do not discuss tumbling in this work.}. This leaves us with three possible IR scenarios.

\subsection*{(I) Conformal fixed point}

 In the first scenario, the theory flows to a conformal field theory (CFT). When the CFT is weakly coupled,  the renormalization group flow from the UV to the IR is very slow. The UV matter content (fermions) can be considered the IR gapless excitations, and the anomalies are automatically matched. Anomaly matching by strongly interacting CFT is still an open problem; we have nothing to say here.  The existence of a well-controlled Banks-Zaks fixed point implies a strictly weakly coupled CFT. However, such a reliable fixed point can only be obtained in the large-$N$ limit. The $3$-loop $\beta$-function reads
\begin{eqnarray}
\beta(g)=&-\beta_0\frac{g^3}{(4\pi)^2}-\beta_1\frac{g^5}{(4\pi)^4}-\beta_2\frac{g^7}{(4\pi)^6}\,.
\label{beta function to 3 loops}
\end{eqnarray}
 If $\beta_0>0$ and $\beta_1<0$, the theory flows to an IR fixed point, $g^2_*=-\frac{(4\pi)^2\beta_0}{\beta_1}$, up to corrections from $\beta_2$. In the large-$N$ limit and close to the boundary of the asymptotic-freedom region, the contribution from the third term is suppressed compared to the second term, and thus, this term and higher-order terms can be safely neglected. Our theories, however, do not admit large-$N$ analysis. Yet, as we shall discuss, in a few cases, the numerical value of the third term is extremely small compared to the first two terms, so one can conclude that such a weakly-coupled CFT exists.  In Appendix \ref{2loop beta function}, we work out the fixed points using the $2$ and $3$-loop $\beta$-functions. We consider values of $\frac{g_*^2}{4\pi}<0.1$, using both the $2$-loop and $3$-loop calculations, small enough to conclude that the theory has an IR fixed point. Also, the perturbative nature of the $\beta$-function calculations will be trusted when the third term in (\ref{beta function to 3 loops}) is small compared to the first two terms. More stringent coupling constant values at the fixed point could also be assumed. This, however, will only mean doing more work to find out the fate of the IR phases of such theories.

\subsection*{(II) Composite massless fermions}

 In the second scenario, the theory becomes strongly coupled; it confines (for $N$ even), preserves the global symmetries, and flows to a phase of composite massless fermions. This can happen in the non-bosonic theories $N=5,6,10$.  We sketch how one can systematically search for such composites. Let ${\cal F}_i$ be a gauge-invariant fermionic operator (a composite that transforms as a left-handed Weyl fermion under the Lorentz group) built of $\psi$ and $\chi$:
 \begin{eqnarray}
 {\cal F}_i= \psi^{\kappa_{_i}} \chi^{\rho_i}\,,
 \end{eqnarray}
where $\kappa_i, \rho_i\in \{0\}\cup \mathbb Z^+$, and we suppressed the color and spinor indices to reduce notational clutter. Insertion of gluon fields can be used whenever fermi statistics cause $ {\cal F}_i$ to vanish.  Using the convention that $\psi$ and $\chi$ carry $2$ and $N-2$ indices, respectively, and demanding that ${\cal F}_i$ be a gauge invariant fermion yields the two conditions:
 \begin{eqnarray}
 2\kappa_i+(N-2)\rho_i \in N\mathbb  Z^+\,, \quad \kappa_i+\rho_i\in (2\mathbb Z^+-1)\,.
 \end{eqnarray} 
 The $U(1)_A$ charge of ${\cal F}_i$ is 
 \begin{eqnarray}
 q_{{\cal F}_i}=\frac{-\kappa_i N_\chi+\rho_i N_\psi}{r}\,.
 \end{eqnarray}
  Generally, the composites ${\cal F}_i$ transform in higher representations of $SU(n_\psi)$ and $SU(n_\chi)$, making the process of matching anomalies containing flavor groups a daunting task. Thus, it is more convenient to start with matching $\left[U(1)_A\right]^3$, $U(1)_A[\mbox{grav}]$, and  nonperturbative ${\mathbb Z}_p^{d\chi}$ anomalies.  For generality, we assume there are ${\cal N}_i$ copies of composites ${\cal F}_i$. Then, matching these anomalies gives the conditions
  \begin{eqnarray}
  \nonumber
  \sum_{i}{\cal N}_i  q_{{\cal F}_i}^3&=&q_\psi^3n_\psi \mbox{dim}_\psi+q_\chi^3n_\chi \mbox{dim}_\chi\,,\\
  \nonumber
  \sum_{i}{\cal N}_i  q_{{\cal F}_i}&=&q_\psi n_\psi \mbox{dim}_\psi+q_\chi n_\chi \mbox{dim}_\chi\,,\\
  \nonumber
 2 \sum_{i}{\cal N}_i&=&2  n_\chi \mbox{dim}_\chi \,(\mbox{mod}\, p)\,,\quad
  (p^2+3p+2) \sum_{i}{\cal N}_i=  (p^2+3p+2)  n_\chi \mbox{dim}_\chi \,(\mbox{mod}\, 6p)\,.\\
  \label{the 3 conditions}
  \end{eqnarray}
 The number of the IR fermionic species ${\cal N}=\sum_{i}{\cal N}_i$ is bounded from above by the a-theorem:
 \begin{eqnarray}
 \underbrace{\underbrace{2(N^2-1)}_{\mbox{gluons}}+\frac{7}{4}\left(n_\psi \mbox{dim}_\psi+n_\chi \mbox{dim}_\chi\right)}_{\mbox{UV degrees of freedom}}\geq \underbrace{\frac{7{\cal N}}{4}}_{\mbox{IR degrees of freedom}}\,.
 \end{eqnarray}
In principle, one could systematically search for copies of composites $\{{\cal N}_1, {\cal N}_2,...\}$ that satisfy (\ref{the 3 conditions}). However, this would require finding the partitions of ${\cal N}$ (all integers that their sums give ${\cal N}$),  a number that grows exponentially with $\sqrt {\cal N}$. The composites that satisfy (\ref{the 3 conditions}) must also match the rest of the anomalies that involve the flavor groups.  In all non-bosonic theories, $N=5,6,10$, we could not find a set of composites that matched the full set of anomalies using the systematic approach sketched above. Simply, the algorithm takes an extremely long time, which makes such a systematic search impractical.

In fact, we can utilize the $\mathbb Z_p^{d\chi}[\mbox{CFU}]$ anomaly to show that in some cases, such candidates, if they exist, cannot solely match this anomaly. This approach was used in \cite{Anber:2019nze} in the case of vector-like theories, and we repeat it here for chiral theories. To this end, we assume that there exists a set of gauge-invariant composite fermions that match $[SU(n_\psi)]^3$, $[SU(n_\chi)]^3$, $[U(1)_A]^3$, $\mathbb Z_p^{d\chi}[U(1)_A]^2$, $\mathbb Z_p^{d\chi}[SU(n_\psi)]^2$, $\mathbb Z_p^{d\chi}[SU(n_\chi)]^2$, $U(1)_A[\mbox{grav}]$, and $\mathbb Z_p^{d\chi}[\mbox{grav}]$ anomalies. Then, we turn the CFU fluxes on $\mathbb M^4$ and perform a $\mathbb Z_p^{d\chi}$ rotation. We denote the UV coefficients that multiply $Q_c$, $Q_\chi$, and $Q_u$ in (\ref{CFU anomaly spin}, \ref{Dirac indices}) by $D_c^{\scriptsize \mbox{UV}}\equiv n_\chi T_\chi$, $D_\chi^{\scriptsize \mbox{UV}}\equiv \mbox{dim}_\chi$, and $D_u^{\scriptsize \mbox{UV}}\equiv q_\chi^2 n_\chi \mbox{dim}_\chi$. Under a discrete chiral rotation, the UV partition function transforms as 
\begin{eqnarray}
 {\cal Z}_{\scriptsize\mbox{UV}}\longrightarrow e^{i\frac{2\pi}{p}\left(D_c^{\scriptsize \mbox{UV}} Q_c+D_\chi^{\scriptsize \mbox{UV}} Q_\chi+D_u^{\scriptsize \mbox{UV}} Q_u\right)} {\cal Z}_{\scriptsize\mbox{UV}}\,,
\end{eqnarray}
 while the IR partition function transforms as\footnote{$Q_c$ does not contribute to the IR phase since the composites are color singlets.}
 \begin{eqnarray}
 {\cal Z}_{\scriptsize\mbox{IR}}\longrightarrow e^{i\frac{2\pi}{p}\left(D_\chi^{\scriptsize \mbox{IR}} Q_\chi+D_u^{\scriptsize \mbox{IR}} Q_u\right)}  {\cal Z}_{\scriptsize\mbox{IR}}\,,
 \end{eqnarray}
where $D_\chi^{\scriptsize \mbox{IR}}, D_u^{\scriptsize \mbox{IR}}\in \mathbb Z$ are group-theoretical coefficients that are chosen to match $\mathbb Z_p^{d\chi}[U(1)_A]^2$ and $\mathbb Z_p^{d\chi}[SU(n_\chi)]^2$ anomalies.  Since the UV-IR anomaly matching is mod $p$, we must have\footnote{$\mathbb Z_p^{d\chi}$ is a good symmetry in the color background, and thus we must have $D_c^{\scriptsize \mbox{UV}}= p\ell_c$.}
\begin{eqnarray}
D_c^{\scriptsize \mbox{UV}}= p\ell_c\,, \quad D_\chi^{\scriptsize \mbox{UV}}- D_\chi^{\scriptsize \mbox{IR}}= p\ell_\chi\,, \quad D_u^{\scriptsize \mbox{UV}}- D_u^{\scriptsize \mbox{IR}}=p\ell_u\,,
\end{eqnarray}
 for some $ \ell_{c,\chi,u} \in \mathbb Z$. Thus, the ratio between  the UV and IR partition functions reads 
\begin{eqnarray}
\frac{ {\cal Z}_{\scriptsize\mbox{UV}}}{ {\cal Z}_{\scriptsize\mbox{IR}}}=e^{i2\pi\left( \ell_c Q_c+\ell_\chi Q_\chi+\ell_u Q_u\right)}\,,
\end{eqnarray}
and the matching of the $\mathbb Z_p^{d\chi}[\mbox{CFU}]$ anomaly requires
\begin{eqnarray}
\ell_c Q_c+\ell_\psi Q_\chi+\ell_u Q_u\in \mathbb Z\,,
\label{condition for CFU}
\end{eqnarray}
for all allowed topological charges. Suppose no integers $\ell_{c,\chi,u}$ exist that satisfy this condition for a given allowed fractional topological charges. In that case, the composites cannot solely match the $\mathbb Z_p^{d\chi}[\mbox{CFU}]$ anomaly. 

A minimal way out would be breaking $\mathbb Z_{p}^{d\chi}\longrightarrow \mathbb Z_{q<p}$ via condensate formation provided that the anomaly  $\mathbb Z_q[\mbox{CFU}]$ vanishes. Usually, a condensate would ordinarily break $SU(n_\psi)$, $SU(n_\chi)$, and $U(1)_A$. Thus, one must postulate that all gauge-invariant condensates charged under these symmetries have zero vacuum expectation values. Otherwise, the condensation of such operators would oversaturate these anomalies, which are assumed to be matched by composites. In addition, one needs to build a neutral operator under the continuous symmetries, charged under $\mathbb Z_{p}^{d\chi}$, and has a non-zero vacuum expectation value. If it exists, such an operator would have a scaling dimension larger than the vanishing lower-order condensates. Although this scenario cannot be ruled out, we find it contrived in the examples of the $2$-index chiral theories we discuss here. 

This leaves us with the possibility that if condition (\ref{condition for CFU}) is violated, the $\mathbb Z_p^{d\chi}[\mbox{CFU}]$ anomaly can be matched by a symmetry-preserving topological quantum field theory (TQFT). In \cite{Cordova:2019bsd,Cordova:2019jqi}, it was shown that the matching of $\mathbb Z_p^{d\chi}$-gravitational anomalies by a unitary and symmetry-preserving TQFT is obstructed on a spin manifold. This obstruction can also be shown to hold in the case of $\mathbb Z_p^{d\chi}$[CFU] anomaly \cite{Anber:2021iip}.

We conclude that if condition (\ref{condition for CFU}) is violated, the theory probably cannot flow to a phase with massless composites.

\subsection*{(III) Spontaneous symmetry-breaking}

 In this scenario, the theory becomes strongly coupled; it confines (for $N$ even) and breaks its global symmetries spontaneously. We say that the theory flows to a spontaneous symmetry-breaking (SSB) phase. An important aspect of this work involves identifying the minimal set of condensates that break global symmetries while matching the anomalies. These condensates break $G^{\scriptsize \mbox{g}}$ down to $H\subset G^{\scriptsize \mbox{g}}$, with the requirement that $H$ remains anomaly-free. Without satisfying this condition, the symmetry breaking alone would not sufficiently match the UV anomaly. It is possible for composite fermions to match a non-vanishing anomaly in $H$, but it is crucial that these fermions do not undermine the matching of the $G^{\scriptsize \mbox{g}}$ anomalies achieved by the condensates. Our focus did not involve searching for composites that could match the anomalous unbroken subgroups.

 Generally, $H$ can be expressed as $H=H^c\times \mathbb Z_{q_1}\times \mathbb Z_{q_2}$, where $H^c$ represents the continuous part of $H$, $\mathbb Z_{q_1}$ ``collectively" denotes the unbroken discrete subgroups of $SU(n_\psi)\times SU(n_\chi)\times U(1)_A$, and $\mathbb Z_{q_2}$ represents the unbroken subgroup of $\mathbb Z_p^{d\chi}$. If the condensates leave a discrete subgroup unbroken, we must examine its anomalies. In addition, if the theory possesses a $ 1$-form/$0$-form mixed anomaly, there can be fractionalization classes, and hence, an ambiguity in calculating the cubic discrete anomalies \cite{Delmastro:2022pfo}. One must ensure the condensates do not leave any discrete anomaly in any fractionalization class\footnote{We would like to thank Erich Poppitz for illuminating discussions about this point.}. In a few examples, we observe that lower-order condensates (such as the $2$-fermion condensates) lead to anomalous unbroken discrete subgroups. Consequently, the formation of other (higher-order) condensates becomes necessary to break the symmetries into non-anomalous subgroups.

In strongly-coupled theories, it is generally believed that higher-order bosonic operators undergo condensation. In this work, through anomaly matching conditions,  we provide kinematical reasons behind this condensation.

Now we turn our attention to the matching of CFU anomalies. Given the better understanding of the nature of the unbroken discrete subgroups of $U(1)_A$ and their anomalies, here, we provide a more in-depth discussion of the CFU anomalies than the earlier work \cite{Anber:2021iip}.  As mentioned above, we encounter two types of such anomalies: $U(1)_A[\mbox{CFU}]$ and $\mathbb Z_p^{d\chi}[\mbox{CFU}]$ anomalies. In all the examples we have examined, we consistently observe the trivialization of $\mathbb Z_p^{d\chi}[\mbox{CFU}]$ by the condensates, which break the $U(1)_A$ symmetry. On the other hand, the matching of the $U(1)_A[\mbox{CFU}]$ anomaly through condensates is a more intricate process that required closer examination.

As emphasized earlier, the color topological charge does not play a role in this particular anomaly. Consequently, we can view it as an anomaly of $U(1)_A$ in the presence of the flavor center and $U(1)_A$ fluxes. In the examples we have studied, the condensates break the flavor center, rendering this anomaly irrelevant. In simpler terms, the full breaking of the flavor center automatically matches the $U(1)_A[\mbox{CFU}]$ anomaly. This can be understood through the following principle: if a triangle (anomaly) diagram involves three abelian symmetries, namely $G_1$, $G_2$, and $G_3$ (in this case, $G_1$ through $G_3$ are the abelian discrete groups corresponding to turning on the CFU fluxes), the complete breaking of at least one of these symmetry groups will resolve the anomaly.

To extract more valuable insights from this anomaly, we can focus our attention solely on the color-$U(1)_A$ fluxes by deactivating the flavor background. In doing so, the $U(1)_A[\mbox{CU}]$ anomaly becomes a mixed anomaly of $U(1)_A$ in the presence of fractional $U(1)_A$ flux (keeping in mind that the $U(1)_A$ flux still needs to combine with the color flux to satisfy the cocycle conditions. Yet, the color topological charge remains uninvolved in the anomaly). Superficially, one might consider this to be equivalent to the $[U(1)_A]^3$ anomaly. However, this is not the case since the latter anomaly only encompasses integer fluxes of $U(1)_A$, whereas the $U(1)_A[\mbox{CU}]$ anomaly incorporates the minimal flux of $U(1)_A$. Consequently, the latter is more restrictive in nature compared to the $[U(1)_A]^3$ anomaly.
Let the discrete flux of $U(1)_A$ be a $\mathbb Z_n$ flux, and let a particular condensate break $U(1)_A$ to $\mathbb Z_m\supseteq \mathbb Z_n$. Then the $U(1)[\mbox{CU}]$ anomaly can be thought of as a $\mathbb Z_m[\mathbb Z_n]^2$ anomaly, which can be checked via (\ref{Zm Zn square anomaly}). If $\mathbb Z_m\subset \mathbb Z_n$ and the anomaly $[\mathbb Z_m]^3$ vanishes, the breaking of $U(1)_A$ to $\mathbb Z_m$ automatically matches the  $U(1)[\mbox{CU}]$ anomaly as the symmetry corresponding to the discrete flux of $U(1)_A$ is broken.
 
Next, we discuss the condensates that cause the symmetries to break. A gauge-invariant condensate is a bosonic operator
 \begin{eqnarray}
 {\cal C}=\psi^{\alpha_\psi}\chi^{\alpha_\chi}\,,
 \end{eqnarray}
 and $\alpha_\psi$ and $\alpha_\chi$ satisfy the conditions
 \begin{eqnarray}
 2\alpha_\psi+(N-2)\alpha_\chi\in N \mathbb Z^+\,, \quad \alpha_\psi+\alpha_\chi\in 2\mathbb Z^+\,.
 \end{eqnarray}
 One needs as many condensates as necessary to break $G^{\scriptsize \mbox{g}}$ to an anomaly-free subgroup. Distinct condensates will break $G^{\scriptsize \mbox{g}}$ down to $H_1\subset G^{\scriptsize \mbox{g}}$,  $H_2\subset G^{\scriptsize \mbox{g}}$, etc. If the subgroups $\{H_1, H_2,..\}$ do not share common generators, $G^{\scriptsize \mbox{g}}$ will break to unity. Finding the breaking patterns of a group $G^{\scriptsize \mbox{g}}$ because of the condensation of single or many operators transforming in the defining or higher-dimensional representations of $G^{\scriptsize \mbox{g}}$ is, in general, a complicated problem. Only a few cases have been discussed in the literature; see, e.g., \cite{Li:1973mq,Elias:1975yd,Wu:1981eb, Ruegg:1980gf, Jetzer:1983ij} and references therein. The question then is, how many condensates does the theory develop in the IR? There is no known answer to this question. However, there must be at least as many condensates as needed to match all anomalies.

\subsection{Minimizing the IR degrees of freedom}

 Beyond 't Hooft anomalies, are there additional sources of information that can be harnessed to make conjectures about the infrared (IR) phase of a strongly coupled theory? In \cite{Appelquist:1999hr,Appelquist:1999vs,Appelquist:2000qg}, a constraint on the structure of strongly coupled asymptotically-free field theories was proposed. The
constraint is an inequality favoring an IR phase with fewer degrees of freedom (DOF). It was also proposed to use the free energy to characterize DOF. The effective degrees of freedom ${\cal A}$ of free $n_B$ massless real scalars and free $n_f$ massless Weyl fermions are given in terms of the free energy density $F$ as ($T$ is an infinitesmal temperature)
 \begin{eqnarray}
 {\cal A}\equiv \frac{90 F}{\pi^2 T^4}=n_B+\frac{7}{4}n_f\,.
 \label{NDOF}
 \end{eqnarray}

First, we may use (\ref{NDOF}) to favor between a phase of composite fermions or a phase with spontaneous symmetry breaking (SSB). As we pointed out above, we could not find composite fermions that matched the anomalies. Yet, one may be tempted to use (\ref{NDOF}) to predict whether the theory flows to an IR CFT. In a weakly-coupled CFT, the IR DOF are the same UV DOF. On the other hand, the DOF in a spontaneously broken phase, assuming the global symmetry $SU(N_f)$ entirely breaks, are $N_f^2-1$ Goldstones\footnote{Notice that a theory that fully breaks its global symmetries will match its 't Hooft anomalies in the IR. We assume that enough condensates form to obey the matching conditions.}. Let us define  $\Delta {\cal A}$ as the difference between the DOF in the two scenarios. Then, we have 
 \begin{eqnarray}
 \Delta {\cal A}=\underbrace{n_\psi^2+n_\chi^2-2}_{\mbox{Goldstones}}-\left\{\underbrace{2(N^2-1)}_{\mbox{gluons DOF}}+\frac{7}{4} \left(n_\psi \mbox{dim}_\psi +n_\chi\mbox{dim}_\chi\right)\right\}\,.
 \end{eqnarray}
 According to the conjecture, a theory with $\Delta {\cal A}>0$ disfavors an SSB phase. It can be easily checked that all the theories in Table \ref{tab:2-index_chiral_theory} yield $\Delta {\cal A}<0$, favoring a phase with broken symmetries. This is to be expected since $n_\psi, n_\chi \sim N$ and $\mbox{dim}_\psi, \mbox{dim}_\chi\sim N^2$. Thus, while the SSB phase has $\sim N^2$ DOF, a phase with CFT has $\sim N^3$ DOF. Then, one may naively conclude that all theories in Table \ref{tab:2-index_chiral_theory} will break their symmetries and flow to a Goldstone phase. This conclusion, however, totally ignores the dynamics of the theory on the way from UV to IR. A theory must enter a strongly-coupled regime to form condensates and break its continuous symmetries, i.e., breaking the symmetries has to happen dynamically since no elementary scalars exist. As we argued above, some of our theories have robust IR fixed points at weak coupling, indicating that it is most unlikely they can form condensates. Consequently, in the subsequent analysis, we avoid employing the aforementioned hypothesis to favor between an SSB or CFT phase. Instead, we use the $\beta$-function analysis to check whether a theory flows to an IR CFT\footnote{This method was used in \cite{Anber:2023urd} to predict the IR phase of a theory with a single adjoint and $N_f$ fundamental flavors of Weyl fermions. It was found that the $\Delta{\cal A}$ calculations are consistent with the prediction of perturbative $\beta$-function. The fact that this analysis does not hold for the $2$-index chiral theories is attributed to the large number of degrees of freedom of a CFT, which always exceeds the number of degrees of freedom of an SSB phase.}.
 
However, assuming the existence of multiple sets of condensates, each capable of accounting for all observed anomalies via SSB, we can employ the aforementioned line of reasoning to make a prediction. Presumably, the set of condensates that causes the flavor group to break into the largest subgroup will be preferred due to its associated reduction in the number of infrared degrees of freedom. 

 The following sections are devoted to systematically applying the above ideas to the concrete theories in Table \ref{tab:2-index_chiral_theory}.  We start our discussion by working out all the details. As we progress through the list of theories, we build on the previous experience and shorten our discussion.

\section{Fermionic theories}
\label{Fermionic theories}

This section systematically studies theories that admit fermionic operators in their spectrum. These are $(N=5, k=1)$, $(N=6, k=2)$, $(N=6,k=1)$, and $(N=10, k=2)$. Our analysis indicates that the first two theories form condensates and break their global symmetries, while the last two flow to a CFT.

\subsection{\fbox{$SU(5), k=1$}}

This theory admits a single Weyl fermion $\psi$ and $n_\chi=9$ flavors of $\chi$ Weyl fermions. In addition, we have $r=\mbox{gcd}(n_\psi T_\psi, n_\chi T_\chi)=\mbox{gcd}(7,27)=1$, indicating that the theory does not possess a discrete chiral symmetry. The solutions to the cocycle conditions (\ref{consistency conditions}) give $\mathbb Z_5 \times \mathbb Z_9$ as the discrete division group.  Thus, the faithful global symmetry is
\begin{eqnarray}
G^{\scriptsize\mbox{g}}=\frac{SU(9)_\chi\times U(1)_A}{\mathbb Z_5 \times \mathbb Z_9}\,,
\label{G global for SU5}
\end{eqnarray}
and the $U(1)_A$ charges of $\psi$ and $\chi$ are
\begin{eqnarray}
q_\psi=-27\,, \quad q_\chi=7\,.
\end{eqnarray}
Since both $q_\psi$ and $q_\chi$ are odd, $(-1)^F\equiv \mathbb Z_2^F$ fermion-number symmetry, which acts on $(\psi,\chi)$ as $(\psi,\chi)\longrightarrow -(\psi,\chi)$, is a subgroup of $U(1)_A$. 

The topological charges of the CFU fluxes are given by:
\begin{gather}
    Q_{c} =  \f{4m^{2}}{5}\,, m \in \Z_{5}\,, \quad
    Q_{\chi} = \f{8p'^{2}}{9}\,,p' \in \Z_{9}\,,\quad 
    Q_{u} = s^{2}\,, s \in \mathbb Z_{45}\,,
\end{gather}
and  $(m, p',s)$ are chosen to satisfy (\ref{consistency conditions}). 
The theory admits a set of anomalies listed in Table \ref{tab:anomalies-su5} (from here on, we give the phase of the corresponding anomaly).
\begin{table}[h]
\centering
\begin{tabular}{|l|l|l|}
\hline
Anomaly & Equation & Value \\
\hline
$[U(1)_{A}]^{3}$ & $\kappa_u^3=n_{\psi} q_{\psi}^{3} \dim \psi + n_{\chi} q_{\chi}^{3} \dim \chi$ & $-264375$ \\
$U(1)_{A} [SU(9)_{\chi}]^{2}$ & $q_{\chi} \dim \chi$ & $70$ \\
$\left[SU(9)_\chi\right]^3$ & $\dim \chi$ & $10$ \\
$U(1)_{A} [\text{grav}]$ & $2(n_{\psi} q_{\psi} \dim \psi + n_{\chi} q_{\chi} \dim \chi)$ & $450$ \\
$U(1)_{A} [\text{CFU}]$ & $q_{\chi} \dim \chi \, Q_{\chi} + \k_{u^{3}} Q_{u}$ & $\frac{560}{9}p'^2-264375s^2$ \\
\hline
\end{tabular}
\caption{Anomalies of $SU(5), k=1$.}
\label{tab:anomalies-su5}
\end{table}

Notice that, as pointed out above, the $U(1)_{A} [\text{CFU}]$ anomaly does not depend on the color topological charge. 
We can also put the theory on $\mathbb {CP}^2$ by employing fluxes in the centers of $SU(5)$ and $SU(9)_\chi$ accompanied by a $U(1)_A$ flux, as can be easily checked from (\ref{consistency conditions nonspin}).

The $2$-loop and $3$-loop $\beta$-function analysis show that the theory has an IR fixed point at {\em somewhat large} coupling-constant: $\frac{g_*^2}{4\pi}\approx 0.64$ and $\frac{g_*^2}{4\pi}\approx 0.34$, respectively. Therefore, such a fixed point is not robust. We conclude that either the theory forms composite fermions or flows to an SSB phase. 


\subsection*{Matching by composites} 

We used the systematic approach discussed in Section \ref{Anomaly matching and the IR phase} to search for a set of composite fermions. We found a pair of operators 
\begin{eqnarray}
{\cal F}_1=\psi \chi^6\,, \quad {\cal F}_2=\psi^7\chi^{22}\,,
\end{eqnarray}
with ${\cal N}_1=36$ and ${\cal N}_2=9$ copies that matched the $\left[U(1)_A\right]^3$ and $U(1)_A[\mbox{grav}]$ anomalies. Yet,  this pair failed to match the $U(1)_A\left[SU(9)_\psi\right]^2$ anomaly. The upper bound on the number of the IR fermion species is ${\cal N}\sim 132$. The large number of partitions of ${\cal N}$ is  ${\cal O}(10^7)$, which hindered the abilities of our search algorithm. We failed to find a set of composites that matches the full set of anomalies. 

\subsection*{Matching by condensates} 

We now turn to the formation of condensates. The lowest-order condensate is 
\begin{eqnarray}
{\cal C}_1^i=\epsilon^{a_1a_2a_3a_4a_5}\epsilon_{\alpha_1\alpha_2} \psi_{(a_1a_2)}^{\alpha_1}\chi_{[a_3a_4a_5]}^{\alpha_2,\,i}\,,
\end{eqnarray}
where $a_1,.., a_5$ are color, $\alpha_1, \alpha_2$ are spinor, and $i$ is a $SU(9)_\chi$ flavor indices. This condensate vanishes identically owing to the symmetrizing over $a_1, a_2$. Yet, one can evade this problem by inserting gauge-covariant gluonic fields $(f_{\mu\nu}^c)_{a_i}^{a_j}\sigma^{\mu\nu}$:
\begin{eqnarray}
{\cal C}_1^i\longrightarrow \tilde{\cal C}_1^i= \epsilon^{a_1a_2a_3a_4a_5}\epsilon_{\alpha_1\alpha_2} (f_{\mu\nu}^c)_{a_2}^{a_6}\sigma^{\mu\nu} \psi_{(a_1a_6)}^{\alpha_1}\chi_{[a_3a_4a_5]}^{\alpha_2,\,i}\,.
\end{eqnarray}
This trick will always be followed whenever the statistics of indices cause some operator to vanish. ${\cal C}_1^i$ transforms in the defining representation of $SU(9)_\chi$, and thus, it breaks it down to $SU(8)$. However, the condensation of ${\cal C}_1^i$ leaves a $U(1)$ generator of $SU(9)\times U(1)_A$ unbroken. To see that, we go to a basis where ${\cal C}_1^i\propto \delta_{i,9}$. In this basis, the unbroken $SU(8)$ group acts on the $8\times 8$ upper block matrices of the original $9\times 9$ unitary matrices of $SU(9)$. Now, it is easy to see that the $SU(9)$ Cartan generator $ H_8=\mbox{diag}\left(1,1,...,-8\right)$ combines with the $U(1)_A$ generator to leave the vacuum $ \delta_{i,9}$ invariant:
\begin{eqnarray}
e^{i 2\pi (-20\beta) }\left[\begin{array}{ccc} e^{i 2\pi \alpha}&0...& 0\\ ... &...&...\\ 0&...& e^{i 2\pi(-8 \alpha)}\end{array}\right] \left[\begin{array}{c} 0\\...\\1\end{array}\right]=\left[\begin{array}{c} 0\\...\\1\end{array}\right]\,,
\label{the breaking of SU9}
\end{eqnarray}
where $\alpha$ and $\beta$ are the Cartan and $U(1)_A$ phases, respectively.
The direction $2\alpha=-5\beta$ is the unbroken $U(1)$ direction. The unbroken $SU(8)$ has a non-vanishing cubic anomaly. In addition, the unbroken $U(1)$ symmetry inherits the $U(1)_A[\mbox{grav}]$ anomaly, signaling that such breaking is incomplete or inconsistent with the anomaly-matching conditions. 

Another condensate is (we suppress color and spinor indices to reduce clutter)
\begin{eqnarray}
{\cal C}_2^{(ij)}= \psi^2\chi^{(i}\chi^{j)}\,,
\end{eqnarray}
which transforms in the $2$-index symmetric representation of the flavor group\footnote{We also insert gluons if the statistics cause the condensate to vanish.}. The general form of the ``Higgs" potential of the condensate is
\begin{eqnarray}
V({\cal C}_2) = -\f{1}{2} \m^{2} {\cal C}_2^{(ij)} {\cal C}_{2(ij)} + \f{1}{4} \l_{1} ({\cal C}_2^{(ij)} {\cal C}_{2(ij)})^{2} + \f{1}{4} \l_{2} ({\cal C}_{2(ij)} {\cal C}_2^{(jk)} {\cal C}_{2(kl)} {\cal C}^{(li)}_2)\,,
\end{eqnarray}
for some real parameters $\mu^2>0$, $\lambda_1$, and $\lambda_2$. In the case $\l_{2} > 0$, the condensate has a non-zero vacuum expectation value and we can pick the form of the condensate to be ${\cal C}_2 \propto I_{9}$ \cite{Li:1973mq}. This breaks $SU(9)$ to the anomaly-free subgroup $SO(9)$. 

Is there a combination of $SU(9)_{\chi} \times U(1)_{A}$ that breaks to a remaining $U(1)$ symmetry? As $U(1)_{A}$ is abelian, we need to consider the subgroup generated by the Cartan subalgebra of $SU(9)$. The ``unnormalized" generators of the Cartan subalgebra of $SU(9)$ are:
\begin{eqnarray}
[H_m]_{ij}=\sum_{k=1}^m\delta_{ik}\delta_{jk}-m\delta_{i,m+1}\delta_{j,m+1}\,,\quad m=1,2,...,8\,.
\end{eqnarray}
A general $SU(9)$ element generated by the Cartan subalgebra has the form $\exp(2\pi i \a_{m} H_{m})$, $m=1,2,..,8$ and $\alpha_m \in \mathbb [0,1)$. A combined $SU(9) \times U(1)_{A}$ transformation acts on ${\cal C}^{ij}$ via:
\begin{eqnarray}
{\cal C}'^{(ij)}_2 = e^{2\pi i ( -40\b)} \left(e^{2\pi i \a_{m} H_{m}} \right)^{ik} \left(e^{2\pi i \a_{m} H_{m}} \right)^{jl} {\cal C}_{2(kl)}\,,
\end{eqnarray}
and should leave the vacuum expectation value invariant. Thus, we need
\begin{eqnarray}
 e^{2\pi i(-40 \b)} \left(e^{2\pi i \a_{m} H_{m}} \right)^{ik} \left(e^{2\pi i \a_{m} H_{m}} \right)^{jl} I_9=I_9\,.
\end{eqnarray}
It can be easily checked that there are no nontrivial solutions to the above equation, indicating that no $U(1)$ direction is left unbroken.

Under the action of $U(1)_{A}$, the condensate transforms as 
$
{\cal C}^{(ij)}_2 = \psi^{2} \chi^{(i} \chi^{j)} \longrightarrow {\cal C}'^{(ij)}_2 = e^{i2\pi( -40\beta)} \psi^{2} \chi^{(i} \chi^{j)}\,,
$
where $\beta \in [0,1)$ is the $U(1)_{A}$ parameter. So it appears that the condensate is invariant under a discrete $\Z_{40}$ subgroup of $U(1)_{A}$. But recall that the global symmetry group includes a division by the $\Z_{5}$ center of the color group.  $\Z_{5}$ is not a subgroup of $SU(9)$, therefore it can only quotient $U(1)_{A}$, so the parameter $\beta$ is in fact a $U(1)_{A}/\Z_{5}$ parameter and $\beta \in [0, 1/5)$. Therefore the condensate only exhibits an unbroken $\mathbb Z_{8}$ symmetry.

The discrete symmetry $\mathbb Z_8$ has non-perturbative anomalies, as can easily be checked using (\ref{nonperturbative discrete anomaly}) and (\ref{twisted nonperturbative discrete anomaly}), meaning that the condensation of  ${\cal C}_2^{(ij)}$ is insufficient to match the full set of anomalies. Notice that since both $\psi$ and $\chi$ have odd charges under $U(1)_A$, any unbroken discrete subgroup of $U(1)_A$ necessarily contains $(-1)^F$, and thus, we can use the twisted group $\mbox{Spin}^{\mathbb Z_{2m}}$ to detect the nonperturbative anomaly as given from (\ref{twisted nonperturbative discrete anomaly}).  Moreover, since the theory does not possess a $1$-form symmetry, there is no ambiguity in calculating the discrete-symmetry anomaly \cite{Delmastro:2022pfo}. 

In searching for a condensate that does not leave behind a non-anomalous $U(1)$ or discrete subgroup, we consider the most general bosonic operator:
\begin{eqnarray}
{\cal C}=\psi^{\alpha_\psi}\chi^{\alpha_\chi}\,, \quad 2\alpha_\psi+3\alpha_\chi\in 5\mathbb Z^+\,, \alpha_\psi+\alpha_\chi\in 2 \mathbb Z^+\,.
\end{eqnarray} 
This condensate carries a charge of $-27\alpha_\psi+7\alpha_\chi$ under $U(1)_A$, and thus, breaks $U(1)_A$ down to $\mathbb Z_{(-27\alpha_\psi+7\alpha_\chi)/5}$. We used both  (\ref{nonperturbative discrete anomaly}) and (\ref{twisted nonperturbative discrete anomaly}) to check the nonperturbative anomalies of $\mathbb Z_{(-27\alpha_\psi+7\alpha_\chi)/5}$ and found that both untwisted and twisted backgrounds yield the same results.  The lowest-dimensional condensate that breaks $U(1)_A$ to a non-anomalous subgroup has $\alpha_\psi=1$ and $\alpha_\chi=11$. In this case, $\mathbb Z_{10}$ is the anomaly-free subgroup.  

The condensate \begin{eqnarray}{\cal C}_3=\psi\chi^{11}\end{eqnarray} transforms in a higher representation of $SU(9)_\chi$. One can contract $9$ out of the $11$ flavor indices of ${\cal C}_3$ with the Levi-Civita tensor, leaving $2$ free indices. Then, we can rearrange the free indices (possibly with insertions of gluons in case the statistics cause the condensate to vanish) such that  ${\cal C}_3$  transforms in the $2$-index symmetric representation of $SU(9)$:
\begin{eqnarray}
{\cal C}_3^{(ij)}=\psi\chi^{9}\chi^{(i}\chi^{j)}\,.
\end{eqnarray}
  The condensing of ${\cal C}_3^{(ij)}$ breaks $\frac{SU(9)_\chi\times U(1)_A}{\mathbb Z_5 \times \mathbb Z_9}$ down to the anomaly-free subgroup $SO(9)\times \mathbb Z_{10}$. 

Alternatively, one can search for a companion condensate to ${\cal C}_2^{(ij)}$  that breaks $U(1)_A$ to a discrete subgroup $\mathbb Z_q$, such that $\mbox{gcd}(q,8)=2$. This ensures that the formation of these two condensates breaks $U(1)_A$ down to the anomaly-free subgroup $\mathbb Z_2^F$, which is the fermion number. The companion condensate with the lowest dimension is ${\cal C}_4=\chi^{10}$, which, superficially, breaks $U(1)_A$ down to $\mathbb Z_q=\mathbb Z_{14}$. This, however, is an immature conclusion.  One can contract $9$ flavor indices of ${\cal C}_4$ with a Levi-Civita tensor leaving one free index. Then, ${\cal C}_4$ transforms in the fundamental representation of $SU(9)$, and according to the discussion preceding (\ref{the breaking of SU9}), it breaks it down to $SU(8)\times U(1)$.  Because of the unbroken $U(1)$ generator, the condensation of ${\cal C}_2$  along with ${\cal C}_4$ break $\frac{SU(9)_\chi\times U(1)_A}{\mathbb Z_5 \times \mathbb Z_9}$ down to a subgroup that contains the anomalous  $\mathbb Z_{8}$. More than this is needed to match the full set of anomalies.

 We might continue searching for suitable condensates that break $G^{\scriptsize \mbox{g}}$ to an anomaly-free subgroup. However, the lesson from the above discussion is that it is generically a complex exercise. 

Since $SO(9)$ is the largest anomaly-free subgroup of $SU(9)$, the condensation of ${\cal C}_3^{(ij)}$ leads to the smallest number of the IR Goldstones, and hence, we predict that the theory will flow to a phase with the global symmetry broken down to the anomaly-free $SO(9)\times \mathbb Z_{10}$. This is the minimal scenario. However, because of strong dynamics, nothing forbids the theory from forming all kinds of condensates, breaking $G^{\scriptsize \mbox{g}}$ down to the anomaly-free $\mathbb Z_2^F$ fermion number symmetry.

In an equally alternative scenario, $SU(9)$ could be broken down to the anomaly-free $Sp(8)$ by a condensate transforming in the $2$-index anti-symmetric representation. However, since the dimensions of $Sp(8)$ and $SO(9)$ are identical, anomalies and the argument of the number of Godstones cannot distinguish between the two possible symmetry-breaking scenarios. In general, a condensate transforming in the $2$-index symmetric representation of $SU(2N+1)$ breaks this group down to $SO(2N+1)$, while a condensate in the $2$-index anti-symmetric representation breaks it down to $Sp(2N)$. Both $SO(2N+1)$ and $Sp(2N)$ have dimension $N(2N+1)$.

Interestingly, the operator ${\cal C}_3^{(ij)}$ has a scaling dimension of at least $15$ (it could have a higher dimension if gluon fields are needed to avoid the vanishing of the condensate because of fermi-statistics). That such condensate with a large-scaling dimension must condense in the IR to match the complete set of anomalies is remarkable. Generally,  it is natural to expect that a strongly coupled theory forms higher-order condensates. In this example, however, this formation is not a question about the dynamics; rather, it is a necessary condition for the theory to obey the kinematical constraints imposed by anomalies.

Does our proposed condensate ${\cal C}_3^{(ij)}$ match the $U(1)_A[\mbox{CFU}]$ anomaly? The answer is affirmative.  ${\cal C}_3^{(ij)}$ breaks  $SU(9)_\chi$ flavor group down to $SO(9)$. The latter does not have a center symmetry, while the former group has a $\mathbb Z_9$ center. Thus, we conclude that the condensate breaks $\mathbb Z_9$ maximally, matching the $U(1)_A[\mbox{CFU}]$ anomaly. Next, we may turn off the flavor background, restricting ourselves to the color center and $U(1)_A$ (CU) fluxes. In this case, we have $(m,p',s)=(1,0,\frac{1}{5})$, and keeping in mind that the CU anomaly does not depend on the color topological charge, we find that this is an anomaly of the axial current in the background of a $\mathbb Z_5$ flux. The condensation of ${\cal C}_3$ breaks $U(1)_A$ to $\mathbb Z_{10}$. Thus, the  $U(1)_A[\mbox{CU}]$ anomaly becomes the $\mathbb Z_{10}[\mathbb Z_5]^2$ anomaly discussed around Eq. (\ref{Zm Zn square anomaly}). However, from the last line in Table \ref{tab:anomalies-su5}, the anomaly coefficient becomes, $\f{-264375}{5^2} = -10575$, which is $0$ modulo $5$.  Therefore, in this case, the anomaly becomes trivial, and the $U(1)_A[\mbox{CU}]$ anomaly is automatically matched.

\subsection{\fbox{$SU(6), k=2$}}

This theory has a single $\psi$ Weyl fermion along with $5$ flavors of $\chi$ fermions. Thus, the continuous global symmetry is $SU(5)_\chi\times U(1)_A$. The charges  of $\psi$ and $\chi$ under $U(1)_A$ are 
\begin{eqnarray}
q_\psi=-5\,,\quad q_\chi=2\,.
\end{eqnarray}
 Owing to the fact $r=\mbox{gcd}(n_\psi T_\psi, n_\chi T_\chi)=\mbox{gcd}(8,20)=4$, the theory is also endowed with a $\mathbb Z_4^{d\chi}$ chiral symmetry, which is taken to act on $\chi$ with a unit charge. It can be checked that this is a genuine symmetry since neither $\mathbb Z_4$ nor a subgroup of it can be absorbed in rotations in the centers of $SU(6)\times SU(5)_\chi$. To show that, we try to absorb the elements $e^{i\frac{2\pi \ell}{4}}$, $\ell=1,2,3$, in the centers of $SU(6)\times SU(5)_\chi$:
\begin{eqnarray}
\mathbb Z_4: \quad \psi\longrightarrow e^{2\pi i \f{2m}{6}} \psi=\psi\,,\quad 
   \chi\longrightarrow e^{-2\pi i \f{2m}{6}} e^{-2\pi i \f{p'}{5}}\chi= e^{2\pi i \f{l}{4}}\chi\,.
\end{eqnarray}
No values of $m$ and $p'$ satisfy these equations for $\ell=1,2,3$, and therefore, $\mathbb Z_4^{d\chi}$ is a genuine symmetry. An identical procedure is employed in the rest of the theories to ascertain the genuineness of discrete chiral symmetries. 

To determine the faithful global symmetry, we must find the quotient group by solving the consistency conditions (\ref{consistency conditions}). This gives $\mathbb Z_3\times \mathbb Z_5$ as the group we divide by. Putting everything together and remembering that the theory possesses a $\mathbb Z_2^{[1]}$ $1$-form center symmetry, we write the faithful global group:
\begin{eqnarray}
G^{\scriptsize\mbox{g}}=\frac{SU(5)_\chi\times U(1)_A}{\mathbb Z_3 \times \mathbb Z_5}\times \mathbb Z_4^{d\chi}\times \mathbb Z_2^{(1)}\,.
\label{G global for SU56 k2}
\end{eqnarray}
The $\mathbb Z_2^F$ fermion number symmetry is contained in the generators of the product group $U(1)_A\times \mathbb Z_4^{d\chi}$ (notice that the $U(1)_A$ charges of $\psi$ and $\chi$ are odd and even, respectively)
\begin{eqnarray}
\nonumber
\mathbb Z_2^F\subset U(1)_A: \psi\longrightarrow -\psi\,,\quad \chi\longrightarrow \chi\,,\\
\mathbb Z_2^F\subset \mathbb Z_4^{d\chi}: \psi\longrightarrow \psi\,,\quad \chi\longrightarrow -\chi\,.
\label{fermion number symmetry}
\end{eqnarray}

The topological charges of the CFU fluxes are given by:
\begin{gather}
    Q_{c} =  \f{5m^{2}}{6}\,, m \in \Z_{3}\,, \quad
    Q_{\chi} = \f{4p'^{2}}{5}\,,p' \in \Z_{5}\,,\quad 
    Q_{u} = s^{2}\,, s\in \mathbb Z_{15}\,,
\end{gather}
and  $(m, p',s)$ are chosen to satisfy (\ref{consistency conditions}). The anomalies of the theory are listed in Table \ref{tab:anomalies-su6-k2}.
\begin{table}[h]
\centering
\begin{tabular}{|l|l|l|}
\hline
Anomaly & Equation & Value \\
\hline
$[U(1)_{A}]^{3}$ & $\k_{u^{3}} = n_{\psi} q_{\psi}^{3} \dim \psi + n_{\chi} q_{\chi}^{3} \dim \chi$ & $-2025$ \\
$U(1)_{A} [SU(5)_{\chi}]^{2}$ & $q_{\chi} \dim \chi$ & $30$ \\
$\left[SU(5)_{\chi}\right]^{3}$ & $\dim \chi$ & $15$ \\
$\Z_{4}^{d\chi} \left[U(1)_{A}\right]^{2}$ & $\k_{zu^{2}} = n_{\chi} q_{\chi}^{2} \dim \chi$ & $300\,\mbox{mod}\, 4$ \\
$\Z_{4}^{d\chi} \left[SU(5)_{\chi}\right]^{2}$ & $\dim \chi$ & $15\, \mbox{mod}\, 4$ \\
$U(1)_{A} [\text{grav}]$ & $2(n_{\psi} q_{\psi} \dim \psi + n_{\chi} q_{\chi} \dim \chi)$ & $90$ \\
$\Z_{4}^{d\chi} [\text{grav}]$ & $2n_{\chi} \dim \chi$ & $150\, \mbox{mod}\, 8$ \\
$[\Z_{4}^{d\chi}]^3$ &  (\ref{nonperturbative discrete anomaly}) & $2250\, \mbox{mod}\,  24$  \\
$U(1)_{A} [\text{CFU}]$ & $q_{\chi} \dim \chi \, Q_{\chi} + \k_{u^{3}} Q_{u}$ & $24p'^{2} - 2025s^{2}$ \\
$\Z_{4}^{d\chi} [\text{CFU}]$ & $n_{\chi} T_{\chi} Q_{c} + \dim \chi \, Q_{\chi} + \k_{zu^{2}} Q_{u^{2}}$ & $\frac{50}{3}m^{2} + 12p'^{2} + 300s^{2}$ \\
\hline
\end{tabular}
\caption{Anomalies of $SU(6), k=2$.}
\label{tab:anomalies-su6-k2}
\end{table}
It is worth noting that both $\Z_{4}^{d\chi} [\text{grav}]$ and $\Z_{4}^{d\chi} [\text{CFU}]$ anomalies give at most a $\mathbb Z_2$ phase. Also, this theory cannot be put on $\mathbb{ CP}^2$, as there are no solutions to the conditions (\ref{consistency conditions nonspin}). 

We first comment on the possibility that the theory flows to a Banks-Zaks fixed point in the IR. The $2$-loop beta function of this theory gives $\f{g_{*}^{2}}{4\pi} \approx 8.5\gg 1$. This value of the coupling constant is too large for perturbation theory to hold. At $3$-loops, we obtain
$
\f{g_{*}^{2}}{4\pi} \approx 0.73\,.
$
Also, this coupling-constant value is large, so we cannot conclude that our theory flows to a conformal fixed point in the IR. In the following, we examine the possibilities of fermion composites and SSB.

\subsection*{Matching by composites}

Here, we follow the argument in Section \ref{Anomaly matching and the IR phase} to show that composite fermions cannot solely match all the UV anomalies. The UV  $\Z_{4}^{d\chi}[\text{CFU}]$ anomaly of this theory is (unlike the $U(1)_{A} [\text{CFU}]$ anomaly, it is important to notice that the color flux contributes to the $\Z_{4}^{d\chi}[\text{CFU}]$ anomaly)
\begin{eqnarray}
n_{\chi} T_{\chi} Q_{c} + \dim \chi \, Q_{\chi} + \k_{zu^{2}} Q_{u^{2}} =\f{50}{3}m^{2} + 12p'^{2} + 300s^{2}\,.
\end{eqnarray}
In the IR, a set of gauge invariant composite fermions would generate the corresponding $\Z_{4}^{d\chi}[\text{CFU}]$ anomaly:
\begin{eqnarray}
D_{\chi}^{\text{IR}} Q_{\chi} + D_{u}^{\text{IR}} Q_{u}
\end{eqnarray}
for integers $D_{\chi}^{\text{IR}}$ and $D_{u}^{\text{IR}}$. The integers $D_{\chi}^{\text{IR}} $ and $ D_{u}^{\text{IR}}$ are group-theoretical coefficients that are assumed to be found by matching all anomalies of continuous symmetries. 
In the presence of a CFU background flux, the ratio between the UV and IR partition functions after undergoing a $\Z_{4}^{d\chi}$ transformation is given by:
\begin{eqnarray}
\f{\mathcal{Z}^{\text{UV}}}{\mathcal{Z}^{\text{IR}}} = e^{\f{i2\pi}{4} \left( \f{50}{3}m^{2} + \left(12 - D_{\chi}^{\text{IR}}\right) p^{2} + \left(300 - D_{u}^{\text{IR}}\right)s^{2}\right)} = e^{\f{i2\pi}{4} \left( \f{50}{3}m^{2} + d_{\chi} p'^{2} + d_{u} s^{2} \right)}\,,
\end{eqnarray}
where $d_{\chi} = 12 - D_{\chi}^{\text{IR}} \in \Z$ and $d_{u} = 300 - D_{u}^{\text{IR}} \in \Z$. If there exists a particular solution $(m, p', s)$ of the consistency conditions (\ref{consistency conditions}) such that no integers $d_{\chi}, d_{u}$ exists such that 
\begin{eqnarray}
\f{50}{3}m^{2} + d_{\chi} p^{2} + d_{u} s^{2} \in 4\Z, 
\label{tag1}
\end{eqnarray}
then we conclude that composite fermions cannot match the $\Z_{4}^{d\chi}[\text{CFU}]$ anomaly.

Consider $(m, p, s) = (1, 0, 2/3)$. This is a solution to the consistency conditions and, therefore, corresponds to a CFU flux. In the presence of this CFU background, the LHS of (\ref{tag1}) becomes 
\begin{eqnarray}
\f{50}{3} + d_{u} \f{4}{9} = \f{150 + 4d_{u}}{9}\,.
\end{eqnarray}
However, $150 + 4d_{u} \equiv 2 \mod 4$ for any $d_{u} \in \Z$. Therefore, we can conclude that for this theory, composite fermions cannot solely match the $\Z_{4}^{d\chi}[\text{CFU}]$ anomaly in the IR. 

\subsection*{Matching by a condensate}

Without composites, the anomalies are matched by spontaneous symmetry breaking via condensates. First, the $2$-fermion condensate cannot match the anomalies, as it breaks $SU(5)_\chi\times U(1)_A$ down to the anomalous subgroup $SU(4)\times U(1)$. Next,  consider the operator
\begin{eqnarray}
{\cal C}^{(ij)} = \psi^{2} \chi^{(i} \chi^{j)}\,,
\end{eqnarray}
where $i, j$ are $SU(5)_{\chi}$ flavor indices, and in particular, this condensate is in the two-index symmetric irrep of $SU(5)_{\chi}$. Thus, the condensation of this operator breaks $SU(5)$ to the anomaly-free subgroup $SO(5)$. 

Under the action of $U(1)_{A}$, the condensate transforms as 
$
{\cal C}^{(ij)} = \psi^{2} \chi^{(i} \chi^{j)} \longrightarrow {\cal C}'^{(ij)} = e^{2\pi ( 6\beta)} \psi^{2} \chi^{(i} \chi^{j)}
$
where $\beta \in [0,1)$ is the $U(1)_{A}$ parameter. So it appears that the condensate is invariant under a discrete $\Z_{6}$ subgroup of $U(1)_{A}$. But recall that the global symmetry group includes a division by the $\Z_{3}$ center of the color group.  $\Z_{3}$ is not a subgroup of $SU(5)$, therefore it can only quotient $U(1)_{A}$, so the parameter $\beta$ is in fact a $U(1)_{A}/\Z_{3}$ parameter and $\beta \in [0, 1/3)$. Therefore the condensate only exhibits an unbroken $\mathbb Z_{2}$ symmetry, which has no global anomaly, and the $U(1)_{A}$ breaks to a non-anomalous subgroup.

The condensate also breaks $\mathbb Z_4^{d\chi}$ down to $\mathbb Z_2$, leading to $2$ vacua connected via a domain wall. Recalling that the $\Z_{4}^{d\chi} [\text{grav}]$ anomaly is only a $\mathbb Z_2$ phase, the unbroken subgroup $\mathbb Z_2\subset \mathbb Z_4^{d\chi}$ is anomaly free (remember that $\mathbb Z_2$ is also free from nonperturbative anomalies). In addition, the $\mathbb Z_4^{d\chi}[\mbox{CFU}]$ anomaly is valued in $\mathbb Z_2$, meaning that the same condensate saturates it.  The breaking of  $\mathbb Z_4^{d\chi}$ down to $\mathbb Z_2$ will also automatically match the $[\mathbb Z_4^{d\chi}]^3$ anomaly, since $\mathbb Z_2$ is anomaly free. 

Let us examine the fate of the $U(1)_A[\mbox{CFU}]$ anomaly. First, when we turn on the flavor center flux, the breaking of $SU(5)$ into $SO(5)$ matches the anomaly, as the breaking causes the center of $SU(5)$ to break. Next, we solely turn on the color and $U(1)_A$ fluxes. In this case, $s \in \mathbb Z_3$, and the breaking of $U(1)_A$ down to $\mathbb Z_2$ implies that we are after $\mathbb Z_2[\mathbb Z_3]^2$ anomaly. The anomaly coefficient can be read from Table \ref{tab:anomalies-su6-k2}, and according to (\ref{Zm Zn square anomaly}), the anomaly is automatically matched since $\mbox{gcd}(3,2)=1$.

We conclude that the global symmetry $G^{\scriptsize \mbox{g}}$ breaks down to $SO(5)\times\frac{(\mathbb Z_2\subset U(1)_A)\times(\mathbb Z_2\subset  \mathbb Z_4^{d\chi})}{\mathbb Z_2}$. The first $\mathbb Z_2$ symmetry acts only on $\psi$, while the second $\mathbb Z_2$ acts only on $\chi$. Then, from (\ref{fermion number symmetry}), we see that the combination of these symmetries acts like the fermion number, which is left intact in the IR. The extra modding by $\mathbb Z_2$ is employed to avoid overcounting. 

Since $SO(5)$ is the largest anomaly-free subgroup of $SU(5)$, this breaking pattern minimizes the number of Goldstones and is the most favorable scenario.

\subsection{\fbox{$SU(6), k=1$}}

This theory has $2$ flavors of $\psi$ and 10 flavors of $\chi$, and thus, the flavor symmetry is $SU(2)_\psi\times SU(10)_\chi$. The $U(1)_A$ charges are
\begin{eqnarray}
q_\psi=-5\,,\quad q_\chi=2\,.
\end{eqnarray}
Because $r=\mbox{gcd}(n_\chi T_\chi, n_\psi T_\psi)=\mbox{gcd}(40,16)=8$, we may be tempted to conclude the theory has a $\mathbb Z_8$ chiral symmetry. However, one can show that a $\mathbb Z_2$ subgroup of the $\mathbb Z_{10}$ center of $SU(10)_\chi$ can be used to identify elements of $\mathbb Z_8$: 
\begin{eqnarray}
     \chi&:& e^{-2\pi i \f{p'}{10}} e^{2\pi i \f{l}{8}} = e^{2\pi i \f{l'}{8}}\,,
\end{eqnarray}
for $l,l'=1,2,...,7$. For example, setting $p'=-5$ identifies $\ell=1$ and $\ell=5$, etc. In addition, the solutions to the consistency conditions (\ref{consistency conditions}) yield the division group $\mathbb Z_3\times \mathbb Z_2\times \mathbb Z_5$. Thus, the faithful global symmetry is
\begin{eqnarray}
G^{\scriptsize \mbox{g}}=\f{SU(2)_{\psi} \times SU(10)_{\chi} \times U(1)_{A}}{\Z_{3} \times \Z_{2} \times \Z_{5}} \times \Z_{4}^{d\chi} \times \Z_{2}^{(1)}\,.
\end{eqnarray}

The $\beta$-function indicates that the theory flows to an IR fixed point. At $2$ loops, the coupling constant at the fixed point is $\f{g_{*}^2}{4\pi} \approx 0.094$. At $3$ loops, we obtain $\f{g_{*}^2}{4\pi} \approx 0.075$. Both values are much smaller than our threshold value of $0.1$, and the $2$- and $3$-loop analysis is only $10\%$ apart. Also, the $3$-loop to the $2$-loop ratio in (\ref{beta function to 3 loops}) is $\approx 0.2$. Thus, the fixed point is reliable. As we pointed out above, the lowest-order bosonic operator in this theory, $F_{\mu\nu}\sigma^{\mu\nu} \chi\psi$, necessitates the introduction of a color field to prevent its vanishing due to statistics. This is a dimension-$5$ operator, and due to the smallness of the coupling constant, we do not expect this operator to condense. Not to mention that this operator by itself is not enough to match the full set of anomalies, and higher-order condensates must also form to match them. We, thus, conclude that the most probable scenario is that the theory flows to a CFT.

\subsection{\fbox{$SU(10), k=2$}}

The theory admits $3$ flavors of $\psi$ and $7$ flavors of $\chi$. The charges of the fermions under $U(1)_A$ are
\begin{eqnarray}
q_\psi=-14\,,\quad q_\chi=9\,.
\end{eqnarray}
We also have $r = \gcd(N_{\psi}, N_{\chi}) = 4$, so that the theory is endowed with a $\Z^{d\chi}_{4}$ chiral symmetry, which cannot be absorbed in a combination of the centers of the color or flavor groups. 
After solving the consistency equations, we obtain the faithful global symmetry group
\begin{eqnarray}
G^{\scriptsize\mbox{g}} = \f{SU(3)_{\psi} \times SU(7)_{\chi} \times U(1)_{A}}{\Z_{5} \times \Z_{3} \times \Z_{7}} \times \Z^{d\chi}_{4} \times \Z^{(1)}_{2}\,.
\end{eqnarray}

The theory develops a Banks-Zaks fixed point. The $2$ and $3$-loop values of the coupling constant at the fixed point are $\frac{g_*^2}{4\pi^2}\approx0.059$ and $\frac{g_*^2}{4\pi^2}\approx0.046$, respectively. Also, the $3$-loop to the $2$-loop ratio in (\ref{beta function to 3 loops}) is $\approx 0.2$. Thus, like $SU(6), k=1$, this theory is expected to flow to a CFT.

\section{Bosonic theories}
\label{Bosonic theories}

All gauge-invariant operators in this class of theories are bosonic. In the following, we provide a systematic study of this class. 

\subsection{Conformal theories}

We start by listing theories that flow to a conformal fixed point. These theories are displayed in Table \ref{list of bosonic theories}. In each case, the global symmetry is given by
\begin{eqnarray}
G=\frac{SU(n_{\psi})\times SU(n_{\chi})\times U(1)_A}{\Gamma} \times\mathbb Z_p^{d\chi}\times \mathbb Z_2^{(1)}\,.
\end{eqnarray}
We also display the coupling constant $\frac{g_*^2}{4\pi^2}$ at the $2$- and $3$-loop fixed points. The smallness of the coupling constant and its consistency between the $2$- and $3$-loop calculations is an indicator of the robustness of the fixed point. To quantify this robustness, we may truncate the $\beta$-function to the second term in (\ref{beta function to 3 loops}) and find the fixed point is given by $\alpha_*=-\frac{4\pi \beta_0}{\beta_1}$. The existence of such a fixed point implies that the first and second terms possess comparable magnitudes. Consequently, the ratio between the third and second (or first) term $\frac{\alpha_*\beta_2}{4\pi\beta_1}$ represents the error incurred by neglecting the third term. A low ratio indicates the perturbative nature of the fixed point.

The two theories $(N=20, k=4)$ and $(N=44, k=8)$ have the most reliable fixed points. While the theory $(N=28, k=8)$ has $\frac{\alpha_*\beta_2}{4\pi\beta_1}=1.12$, and its fixed point is under question. 
\begin{table}
\begin{center}
    \begin{tabular}{|c|c|c|c|c|c|c|c|c|}\hline
    Theory & $n_{\psi}$ & $n_{\chi}$ & $\Z^{d\chi}_{p}$ & $\Gamma$ & $(q_\psi, q_\chi)$ & 2-loop& 3-loop & $\frac{\alpha_*\beta_2}{4\pi\beta_1}$\\
    \hline
    $SU(16), k=4$ & 3 & 5 & 2 & $\mathbb Z_8 \times \mathbb Z_3 \times \mathbb Z_5 $ & $(-35,27)$ & 0.09 & $0.064$ & $0.62$\\
        \hline
    $SU(20), k=4$ & 4 & 6 & 2 & $\mathbb Z_{10} \times \mathbb Z_4 \times \mathbb Z_3 $ & $(-27,22)$ & $0.017$ & $0.015$ & $0.11$\\
        \hline
    $SU(28), k=8$ & 3 & 4 & 1 & $\mathbb Z_{7} \times \mathbb Z_3 \times \mathbb Z_4 $ & $(-52,45)$  & $0.086$ & $0.051$ & $1.12$\\
        \hline
    $SU(36), k=8$ & 4 & 5 & 1 & $\mathbb Z_{18}\times \mathbb Z_4\times\mathbb Z_5$ & $(-85,76)$ & $0.019$ & $0.016$ & $0.25$\\
        \hline
    $SU(44), k=8$ & 5 & 6 & 1 & $\mathbb Z_{11} \times \mathbb Z_5 \times \mathbb Z_6 $ & $(-126,115)$ & $0.0002$ & $0.0002$ & $0.003$\\
        \hline
    \end{tabular}
   \caption{A list of conformal bosonic theories.}
   \label{list of bosonic theories}
   \end{center}
\end{table}
%

\subsection{Confining theories}

\subsubsection{\fbox{$SU(8), k=4$}}

This theory was studied in \cite{Anber:2021iip}. Here, we revisit it in light of the discrete anomalies not discussed in \cite{Anber:2021iip}. The theory admits $n_\psi=1$ and $n_\chi=3$ flavors of fermions. The fermion charges under $U(1)_A$ are
\begin{eqnarray}
q_\psi=-9\,,\quad q_\chi=5\,.
\end{eqnarray}
Also, the theory admits a $\mathbb Z_2^{d\chi}$ discrete chiral symmetry. Solving the consistency conditions (\ref{consistency conditions}) yield the faithful global symmetry
\begin{eqnarray}
G^{\scriptsize \mbox{g}}=\frac{SU(3)_\chi\times U(1)_A}{\mathbb Z_4\times \mathbb Z_3}\times \mathbb Z_2^{d\chi}\times \mathbb Z_2^{(1)}\,.
\label{global G for  N 8 and k 4}
\end{eqnarray}

The theory admits many anomalies in Table \ref{tab:anomalies-su8 and k4}.
\begin{table}[h]
\centering
\begin{tabular}{|l|l|l|}
\hline
Anomaly & Equation & Value \\
\hline
$\left[SU(3)_\chi\right]^3$ & $\mbox{dim}_\chi$ & 28 \\
$U(1)_A\left[SU(3)_\chi\right]^2$ & $q_\chi\mbox{dim}_\chi$ & 140 \\
$\mathbb Z_2^{d\chi}\left[SU(3)_\chi\right]^2$ & $\mbox{dim}_\chi$ & 28 \\
$\mathbb Z_2^{d\chi}\left[U(1)_A\right]^2$ & $\k_{zu^{2}} = q_\chi^2\mbox{dim}_\chi n_\chi$ & 2100 \\
$U(1)_A[\mbox{grav}]$ & $2(q_\psi\mbox{dim}_\psi+q_\chi\mbox{dim}_\chi n_\chi)$ & 192 \\
$\left[U(1)_A\right]^3$ & $\k_{u^3}=q_\chi^3\mbox{dim}_\chi n_\chi+q_\psi^3\mbox{dim}_\psi$ & -15744 \\
$\mathbb Z_2^{d\chi}[\mbox{grav}]$ & $2\mbox{dim}_\chi n_\chi$ & 168 (trivial) \\
$U(1)_A[\mbox{CFU}]$ &$q_{\chi} \dim \chi \, Q_{\chi} + \k_{u^{3}} Q_{u}$&  $\frac{280p'^{2}}{3}-15744s^2$, $p'\in \mathbb Z_3\,, s\in \mathbb Z_{12}$ \\
$\Z_{2}^{d\chi}[\mbox{CFU}]$ & $n_{\chi} T_{\chi} Q_{c} + \dim \chi \, Q_{\chi} + \k_{zu^{2}} Q_{u}$ & $\f{27m^{2}}{2} + \f{56p'^{2}}{3}+2100s^{2} \,, m \in \Z_{4}$\\
\hline
\end{tabular}
\caption{Anomalies of $SU(8), k=4$.}
\label{tab:anomalies-su8 and k4}
\end{table}
In addition, the theory admits a $\mathbb Z_2^{d\chi}[\mbox{CFU}]$ anomaly, which yields a phase of $\pi$ upon turning on a flux with, e.g.,  $(m,p',s)=(1,0,\frac{1}{4})$, i.e., this is a $\mathbb Z_{4}\subset U(1)_A$ flux. We also find that there is an anomaly of $\mathbb Z_2^{d\chi}$ on a nonspin manifold, as the partition function acquires a phase of $\pi$ by turning on a pure $\mathbb Z_2^{(1)}$ flux on $\mathbb{CP}^2$. 

In \cite{Anber:2021iip}, it was argued that all the anomalies could be matched by condensing two operators: 
\begin{eqnarray}
{\cal C}_1^i=\psi \chi^i\,,\quad {\cal C}_{2}^{i_4}=\epsilon_{i_1 i_2i_3}\chi^{i_1}\chi^{i_2}\chi^{i_3}\chi^{i_4}\,.
\end{eqnarray}
 Let us review the anomaly matching using these two operators and comment on why they cannot match the discrete anomalies. 

Both operators ${\cal C}_1^i$ and ${\cal C}_2^i$ transform in the defining representation of $SU(3)$ and break it down to the anomaly-free $SU(2)$ (it has no Witten anomalies because the dimensions of the representations are even). Yet, the condensation of ${\cal C}_1^i$ or ${\cal C}_2^i$ leaves behind an unbroken $SU(3)$ generator. We take ${\cal C}_1^i\propto \delta_{i,1}$ and ${\cal C}_2^i\propto \delta_{i,1}$ and parametrize the $SU(3)$ matrix that corresponds to the unbroken Cartan generator of $SU(3)$ as $\mbox{diag}\left(e^{i 4\pi \alpha}, e^{-i 2\pi \alpha},e^{-i 2\pi \alpha}\right)$. Then, under $SU(3)_\chi\times U(1)_A\times \mathbb Z_{2}^{d\chi}$, the operators transform as
\begin{eqnarray}
{\cal C}_1^i\longrightarrow e^{i 4\pi \alpha-i 8\pi \beta +i n\pi}{\cal C}_1^i\,, \quad {\cal C}_2^i\longrightarrow e^{i 40 \pi \beta +i 4\pi \alpha}{\cal C}_2^i\,,
\label{C1 transformation}
\end{eqnarray}
where $\beta$ corresponds to the $U(1)_A$ transformation, whereas $n=1$ corresponds to the $\mathbb Z_2^{d\chi}$ transformation. Taking $\alpha=-\frac{5}{24}$, $\beta=\frac{1}{48}$, and $n=1$ leaves ${\cal C}_1^i$ and ${\cal C}_2^i$ invariant under the combined transformations of $SU(3)_\chi\times U(1)_A\times \mathbb Z_{2}^{d\chi}$. This superficially hints at an unbroken $\mathbb Z_{24}$ symmetry. However, owing to the modding by $\mathbb Z_4$ in (\ref{global G for  N 8 and k 4}), the genuine unbroken subgroup is $\mathbb Z_6$. This unbroken symmetry can be written as $\mathbb Z_6=\mathbb Z_2\times \mathbb Z_3$, where $\mathbb Z_3$ is a genuine subgroup of $U(1)_A$. This can be seen by setting $n=0$, then we find that ${\cal C}_1^i$ and ${\cal C}_2^i$ are left invariant by taking $\alpha=-\frac{5}{12}$ and $\beta=\frac{1}{24}$. Remembering the modding by $\mathbb Z_4$ in (\ref{global G for  N 8 and k 4}), we conclude that there is a $\mathbb Z_3$ unbroken subgroup of $U(1)_A$. 

It is straightforward to calculate the $\mathbb Z_3$ anomaly using (\ref{nonperturbative discrete anomaly}) to find that it is non-vanishing, meaning that the condensation of ${\cal C}_1^i$ and ${\cal C}_2^i$ is not enough to match the complete set of anomalies. 
The way out is to consider the condensation of the operator
\begin{eqnarray}
{\cal C}_3^{(ij)}=\psi^2\chi^{(i}\chi^{j)}\,,
\end{eqnarray}
which transforms in the $2$-index symmetric representation of $SU(3)_\chi$ and breaks it down to $SO(3)$. 
$U(1)_{A}$ is broken to $\Z_{2}$, after taking into account the modding by $\mathbb Z_4$ in (\ref{global G for  N 8 and k 4}). The $\mathbb Z_2^{d\chi}[\mbox{CFU}]$ anomaly is automatically matched as $U(1)_A$ is broken down to $\mathbb Z_2$ (remember, however, that this $\mathbb Z_2$ is the fermion number since both fermions carry odd charges under $U(1)_A$, and the fermion number is gauged). Recalling that we had to turn on a $\mathbb Z_4\subset U(1)_A$ flux in the first place to see this anomaly (a $\pi$ phase), the breaking of $U(1)_A$ to a smaller subgroup than $\mathbb Z_4$ (in this case $\mathbb Z_2\subset \mathbb Z_4$) trivializes the anomaly.  Thus, at this level, one does not need to break $\Z_{2}^{d\chi}$. This differs from the findings in \cite{Anber:2021iip}, where it was argued that the CFU anomaly is not trivial. Here, we arrive at a different IR condensate by scrutinizing the discrete subgroups of $U(1)_A$. 

What about matching the anomaly of $\mathbb Z_2^{d\chi}$ on $\mathbb {CP}^2$? Since this anomaly is valued in $\mathbb Z_2$, it can be matched by a TQFT on a nonspin manifold, as was argued in \cite{Cordova:2019bsd}. Yet, another scenario is to form the condensate ${\cal C}_1^i$, which breaks $\mathbb Z_2^{d\chi}$ to unity (remember that the $\mathbb Z_2^{(1)}$ $1$-form symmetry is unbroken assuming confinement). Thus, the condensation of both ${\cal C}_1^i$ and ${\cal C}_3^{(ij)}$ match all anomalies and break the global group down to $SO(3)$, resulting in $2$ vacua (because of the breaking of $\mathbb Z_2^{d\chi}$) connected via domain walls.

\subsubsection{\fbox{$SU(8), k=2$}}

This case was also considered briefly in \cite{Anber:2021iip}. The theory admits $2$ flavors of $\psi$ fermions and $6$ flavors of $\chi$ fermions. The $U(1)_A$ charges of the fermions are
\begin{eqnarray}
q_\psi=-9\,, \quad q_\chi=5\,.
\end{eqnarray}
Since $\mbox{gcd}(N_\psi, N_\chi)= 4$, one may naively conclude that the discrete symmetry is $\mathbb Z_4$. Yet, two elements of $\mathbb Z_4$ are identified with elements in $\mathbb Z_2 \subset \mathbb Z_6$, where $\mathbb Z_6$ is the center of $SU(6)_\chi$. This leaves us with $\mathbb Z_2^{d\chi}$ as the genuine discrete group, which we take to act solely on $\chi$. The faithful global symmetry is 
\begin{eqnarray}
G^{\scriptsize \mbox{g}}=\frac{SU(2)_\psi\times SU(6)_\chi\times U(1)_A }{\mathbb Z_4\times \mathbb Z_6}\times \mathbb Z_{2}^{d\chi}\times \mathbb Z_2^{(1)}\,.
\label{global S  of SU 8 and k=2}
\end{eqnarray}
The UV theory has the `t Hooft anomalies in Table \ref{tab:anomalies-su8 and k2}.
\begin{table}[h]
\centering
\begin{tabular}{|l|l|l|}
\hline
  Anomaly & Equation & Value \\
  \hline
  $[SU(6)_\chi]^3$ & $\text{dim}_\chi$ & $28$ \\
  $U(1)_A[\text{grav}]$ & $2(q_\psi\text{dim}_\psi+q_\chi\text{dim}_\chi n_\chi)$& $384$ \\
  $\mathbb{Z}_2^{d\chi}[\text{grav}]$ & $\text{dim}_\chi n_\chi$ & $336$ (trivial) \\
  $U(1)_A[SU(6)_\chi]^2$ & $q_\chi \text{dim}_\chi$& $140$ \\
  $U(1)_A[SU(2)_\psi]^2$ & $q_\psi\text{dim}_\psi$& $-324$ \\
  $[U(1)_A]^3$ & $q_\psi^3\text{dim}_\psi+q_\chi^3\text{dim}_\chi n_\chi$& $-31488$ \\
  $\mathbb{Z}_{2}^{d\chi}[SU(6)_\chi]^2$ & $\mbox{dim}_\chi$ & $28$ (trivial) \\
  $\mathbb{Z}_{2}^{d\chi}[U(1)_A]^2$ & $q_\psi^2 \mbox{dim}_\psi n_\psi+q_\chi^2 \mbox{dim}_\chi n_\chi$ & $4200$ (trivial) \\
  \hline
\end{tabular}
\caption{Anomalies of $SU(8), k=2$.}
\label{tab:anomalies-su8 and k2}
\end{table}
The $\mathbb Z_2^{d\chi}[\mbox{CFU}]$ anomaly does not provide new information. However, there is a non-trivial $\mathbb Z_2^{d\chi}[\mbox{CFU}]_{\mathbb {CP}^2}$ anomaly (a $\pi$ phase) in the background of a CFU configuration with all fluxes turned on, e.g., $(m, p, p', s) = (1, 1, 1, -5/12)$.

The condensation of the operator
\begin{eqnarray}
{\cal C}_{1\; j}^i=\psi_j \chi^i
\end{eqnarray}
break $SU(2)_\psi\times SU(6)_\chi$ down to $SU(2)\times SU(4)$. The unbroken $SU(4)$ is anomalous. 

Another operator is
\begin{eqnarray}
{\cal C}_2^{[i_1i_2]}=\psi^2 \chi^{[i_1}\chi^{i_2]}\,,
\end{eqnarray}
which is neutral under $SU(2)_\psi \times \mathbb Z_2^{d\chi}$ but transforms in the $2$-index anti-symmetric representation of $SU(6)$ and breaks it down to the anomaly-free $Sp(6)$\footnote{Alternatively, one could propose the formation of a condensate transforming in the $2$-index symmetric representation of $SU(6)$. This condensate, however, would break $SU(6)$ down to $SO(6)$, resulting in a larger number of Goldstones.}. In addition, the condensation of ${\cal C}_2^{[i_1i_2]}$ breaks $U(1)_A$ to the anomaly-free $\mathbb Z_2$, after taking into account the modding by $\mathbb Z_4$ in (\ref{global S  of SU 8 and k=2}). What about the $\mathbb Z_2^{d\chi}[\mbox{CFU}]_{\mathbb {CP}^2}$  anomaly? Remember that one needs to turn on a configuration with $U(1)_A$ flux in $\mathbb Z_{12}$. Since $U(1)_A$ breaks down to $\mathbb Z_2$, the anomaly trivializes. Recall that this $\mathbb Z_2$ is the fermion number since both fermions have odd charges under $U(1)_A$, and that the fermion number is gauged.    Thus, the condensation of ${\cal C}_2^{[i_1i_2]}$ leaves behind the unbroken $\frac{SU(2)_\psi\times Sp(6)}{\mathbb Z_2}\times \mathbb Z_2^{d\chi}$ subgroup and matches all anomalies\footnote{The symplectic group $Sp(2N)$ has a $\mathbb Z_2$ center symmetry, see, e.g., \cite{Anber:2014lba}. This is why we needed to mod by $\mathbb Z_2$ that is common between $Sp(6)$ and $SU(2)_\psi$.}.

\subsubsection{\fbox{$SU(12), k=4$}}

The number of flavors in this case is $n_{\psi} = 2$ and $n_{\chi} = 4$, and the $U(1)_{A}$ charges are:
\begin{eqnarray}
q_{\psi}=-10\,,\quad  q_{\chi}=7\,.
\end{eqnarray}
Since $\mbox{gcd}(n_\chi T_\chi, n_\psi T_\psi)=\mbox{gcd}(40,28)=4$, one may conclude that the theory is endowed with a $\mathbb Z_4$ chiral symmetry that acts on $\chi$.  However, this $\mathbb Z_4$ is the center of the $SU(4)_\chi$ flavor symmetry. Therefore, the theory does not possess a discrete chiral symmetry. Solving the consistency conditions (\ref{consistency conditions}), we find that the faithful global symmetry group is:
\begin{eqnarray}
G^{\scriptsize\mbox{g}} = \f{SU(2)_{\psi} \times SU(4)_{\chi} \times U(1)_{A}}{\Z_{3} \times \Z_{2} \times \Z_{2}}  \times \Z^{(1)}_{2}\,.
\label{global for N 12 k 4}
\end{eqnarray}

This theory is endowed with the anomalies in Table \ref{tab:anomalies-su 12 k 4}. The theory does not possess a Witten anomaly of $SU(2)_\psi$ since $\mbox{dim}_\psi=66$ is an even number.
\begin{table}[h]
\centering
\begin{tabular}{|l|l|l|}
\hline
Anomaly & Equation & Value \\
\hline
$[U(1)_{A}]^{3}$ & $q_\chi^3\mbox{dim}_\chi n_\chi+q_\psi^3 n_\psi\mbox{dim}_\psi$ & -65448 \\
$U(1)_{A} [SU(2)_{\psi}]^{2}$ & $q_\psi \mbox{dim}_\psi$ & -780 \\
$U(1)_{A} [SU(4)_{\chi}]^{2}$ & $q_\chi\mbox{dim}_\chi$ & 462 \\
$[SU(4)_{\chi}]^{3}$ & $\mbox{dim}_\chi$ & 66 \\
$U(1)_{A} [\text{grav}]$ & $ \k_{u^{3}} =q_\chi\mbox{dim}_\chi n_\chi+q_\psi n_\psi\mbox{dim}_\psi$ & 576 \\
$U(1)_{A} [\text{CFU}]$ & $q_{\psi} \dim_\psi \, Q_{\psi}+q_{\chi} \dim_\chi \, Q_{\chi} + \k_{u^{3}} Q_{u}$  & $390 p^{2} + \frac{693}{2} p'^{2} - 65448 s^{2}$, $p, p' \in \mathbb Z_2$ \\
\hline
\end{tabular}
\caption{Anomalies of $SU(12), k=4$.}
\label{tab:anomalies-su 12 k 4}
\end{table}

The $2$-loop and the $3$-loop $\beta$-functions predict fixed points at $\frac{g_*^2}{4\pi}=0.514$ and $0.202$, respectively. Both values are large for the fixed points to be robust.

In searching for candidates that break the symmetries spontaneously, let us study the bilinear condensate:
\begin{eqnarray}
{\cal C}_{j}^{i}=\epsilon^{a_1...a_{12}}\left(f_{\mu\nu}^c\right)_{a_2}^{a_{13}}\sigma^{\mu\nu}\epsilon_{\alpha_1\alpha_2} \psi^{\alpha_1}_{j,\,(a_1a_{13})} \chi^{\alpha_2,\, i}_{[a_3..a_{12}]}\,, \quad j=1,2\,, i=1,2,3,4,
\end{eqnarray}
where, as usual, $a_1,a_2,..$ are color indices, $\alpha_1, \alpha_2$ are spinor indices, while $j$ and $i$ are respectively $SU(2)_\psi$ and $SU(4)_\chi$ indices.
The transformation of ${\cal C}_{j}^{i}$ is noteworthy as it occurs in the fundamental representation of $SU(2)\psi$ and the anti-fundamental representation of $SU(4)\chi$. Consequently, upon condensation, it has the potential to break down $SU(2)_{\psi} \times SU(4)_{\chi}$ to $SU(2)_V\times SU(2)$. This symmetry-breaking pattern can be explained as follows \cite{Li:1973mq}.

To create an invariant potential for the $4\times 2$ matrix ${\cal C}_{j}^{i}$, we define the $4\times 4$ matrix $\Phi^{i,i'}\equiv \sum_{j=1}^2{\cal C}_j^i{\cal C}_j^{i'}$ . By considering the effective potential as a trace over quadratic and quartic terms of $\Phi^{i,i'}$, we might initially assume that we can transform to a basis where $\Phi^{i,i'}$ becomes a non-degenerate diagonal matrix. However, this assumption leads to a contradiction because the $4\times 1$ column vectors in $\Phi^{i,i'}$ are dependent due to the construction of $\Phi^{i,i'}$ from a $4\times 2$ matrix. In other words, $\Phi^{i,i'}$ possesses two zero eigenvalues. Hence, we conclude that we can only transform to a basis that diagonalizes $SU(2){\psi} \times (SU(2)\subset SU(4){\chi})$. This results in the diagonal (vector-like) matrix $SU(2)_V$, while $SU(4-2)=SU(2)\subset SU(4){\chi}$ remains unbroken. Both $SU(2)_V$ and $SU(2)\subset SU(4){\chi}$ are subgroups devoid of anomalies. The potential anomaly, namely the Witten anomaly, does not afflict any of these subgroups. The UV particle content ensures that the number of fermions transforming under $SU(2)\psi$ and $SU(2)\subset SU(4)_\chi$ is $\mbox{dim}{\psi}=78$ and $\mbox{dim}{\chi}=66$, respectively, both of which are even numbers. Therefore, none of these groups can exhibit Witten anomalies.

Moreover, due to the $\mathbb Z_3$ modding in (\ref{global for N 12 k 4}), the axial symmetry $U(1)_{A}/\mathbb Z_3$ identifies a transformation phase $\alpha$ with $\alpha+\frac{2\pi}{3}$. The charge of the condensate ${\cal C}_{j}^{i}$ under $U(1)_A$ is $-3$, leading to the breaking of $U(1)_A$ to unity. Hence, we conclude that the $2$-fermion condensate ${\cal C}_{j}^{i}$ successfully saturates all the anomalies and breaks the global symmetry down to $\frac{SU(2)_V \times (SU(2)\subset SU(4){\chi})}{\mathbb Z_2}$, where we mod by the $\mathbb Z_2$ common center of both groups.

\subsubsection{\fbox{$SU(12), k=8$}}

The number of flavors in this case is $n_{\psi} = 1$ and $n_{\chi} = 2$ and the $U(1)_{A}$ charges are:
\begin{eqnarray}
q_{\psi}=-10\,,\quad q_{\chi} =7\,.
\end{eqnarray}
Given that $r=\mbox{gcd}(n_\psi T_\psi, n_\chi T_\chi)=\mbox{gcd}(14,20)=2$,  we may conclude that the theory has a $\Z_{2}$ discrete chiral symmetry. Yet, one can absorb this $\Z_{2}$ in the center of $SU(2)_\chi$, leaving behind no genuine discrete symmetry. After solving the consistency conditions, we find that the faithful global symmetry group is:
\begin{eqnarray}
G^{\scriptsize \mbox{g}} = \f{SU(2)_{\chi} \times U(1)_{A}}{\Z_{3} \times \Z_{2}}  \times \Z^{(1)}_{2}\,.
\label{faithful global N 12 k 8}
\end{eqnarray}

The theory possesses the anomalies in Table \ref{tab:anomalies-su 12 k 8}.
\begin{table}[h]
\centering
\begin{tabular}{|l|l|l|}
\hline
  Anomaly & Equation & Value \\
  \hline
  $[U(1)_{A}]^{3}$ & $ \k_{u^{3}} =q_\chi^3\text{dim}_\chi n_\chi+q_\psi^3\text{dim}_\psi$&$-32724$ \\
  $U(1)_{A} [SU(2)_{\chi}]^{2}$ &$ q_\chi\text{dim}_\chi$&$462$ \\
  $U(1)_{A} [\text{grav}]$ & $q_\chi\text{dim}_\chi n_\chi+q_\psi\text{dim}_\psi$&$288$ \\
  $U(1)_{A} [\text{CFU}]$ & $q_{\chi} \dim_\chi \, Q_{\chi} + \k_{u^{3}} Q_{u}$ & $231 p'^{2} - 32724s^{2}$, $p' \in \mathbb Z_2$ \\
  \hline
\end{tabular}
\caption{Anomalies of $SU(12), k=8$.}
\label{tab:anomalies-su 12 k 8}
\end{table}
The potential Witten anomaly of $SU(2)_\chi$ is absent because $\mbox{dim}_\chi=66$ is an even number.

The $2$-loop and $3$-loop $\beta$-functions do not predict fixed points, and the theory needs to break its symmetries spontaneously by forming condensates. The operator 
\begin{eqnarray}
{\cal C}_1^i=\psi\chi^i\,,
\end{eqnarray}
where the index $i$ is the $SU(2)_\chi$ flavor, breaks the global symmetry down to $U(1)$. To see that, let us fix the vacuum to be $\left[1\quad 0\right]^T$. Then, if a $U(1)$ generator is left unbroken by the vacuum, one should find a nontrivial solution to 
\begin{eqnarray}
\exp\left[i 2\pi \beta \left[\begin{array}{cc} 1&0\\0&-1\end{array}\right]\right]e^{-i 6\pi \alpha I_{2\times 2}}\left[\begin{array}{c} 1\\0\end{array}\right]=\left[\begin{array}{c} 1\\0\end{array}\right]\,.
\end{eqnarray}
It is easy to check that the solution $\beta=3\alpha$ satisfies the above equation, which is the unbroken $U(1)$ direction.
The unbroken $U(1)$ symmetry inherits the UV mixed $U(1)_A[\mbox{grav}]$ anomaly, and thus, condensing ${\cal C}_1^i$ is not enough to match the anomalies. 

Another operator that can condense is
\begin{eqnarray}
{\cal C}_2=\epsilon_{ij}\psi\psi \chi^i\chi^j\,,
\end{eqnarray}
with possible insertions of gluon fields. The operator ${\cal C}_2$ is singlet under $SU(2)$, but it has a charge $-6$ under $U(1)_A$. Because of the modding by $\mathbb Z_3$ in (\ref{faithful global N 12 k 8}), the condensation of ${\cal C}_2$ breaks $U(1)_A$ down to $\mathbb Z_2$, an anomaly-free subgroup. We conclude that the condensation of ${\cal C}_2$ is enough to match the anomalies,  a scenario with the minimum number of Goldstones.

\subsubsection{\fbox{ $SU(20), k=8$}}

The number of flavors is $n_{\psi} = 2$ and $n_{\chi} = 3$, while the $U(1)_{A}$ charges are:
\begin{eqnarray}
q_{\psi}=-27\,,\quad q_{\chi} =22\,.
\end{eqnarray}
Since $r=\mbox{gcd}(n_\psi T_\psi, n_\chi T_\chi)=(36,66)=2$, we might conclude that the theory possesses a $\Z_{2}$ chiral symmetry.  However, this symmetry can be rotated away in the following way. First, according to our choice, the would-be chiral symmetry acts only on $\chi$. Thus, $(\psi,\chi)\longrightarrow (\psi,-\chi)$ under this $\mathbb Z_2$. Next, we apply a transformation by $(-1)^F$, which sends $(\psi,-\chi)\longrightarrow (-\psi,\chi)$. Finally, we apply another transformation by the center of $SU(2)_\psi$, which sends $(-\psi,\chi)\longrightarrow (\psi,\chi)$. This shows that the theory does not possess a discrete chiral symmetry. Finding the solutions to the consistency conditions, the faithful global symmetry group is:
\begin{eqnarray}
G^{\scriptsize\mbox{g}} = \f{SU(2)_{\psi} \times SU(3)_{\chi} \times U(1)_{A}}{\Z_{5} \times \Z_{2} \times \Z_{3}} \times \Z^{(1)}_{2}\,.
\end{eqnarray}

The anomalies of this theory are given in Table \ref{tab:anomalies-su 20 k 8}.
\begin{table}[h]
\centering
\begin{tabular}{|l|l|l|}
\hline
  Anomaly & Equation & Value \\
  \hline
  $\left[U(1)_{A}\right]^{3}$ & $\k_{u^{3}}=q_\chi^3\text{dim}_\chi n_\chi+q_\psi^3\text{dim}_\psi$&$-2197500$ \\
  $U(1)_{A} \left[SU(2)_{\psi}\right]^{2}$ & $q_\psi n_\psi \text{dim}_\psi$&$-5670$ \\
  $U(1)_{A}\left [SU(3)_{\chi}\right]^{2}$ & $q_\chi n_\chi \text{dim}_\chi$&$4180$ \\
  $\left[SU(3)_{\chi}\right]^{3}$ & $\text{dim}_\chi$&$190$ \\
  $U(1)_{A} \left[\text{grav}\right]$ & $2(q_\chi\text{dim}_\chi n_\chi+q_\psi\text{dim}_\psi)$&$2400$ \\
  $U(1)_{A} [\text{CFU}]$ & $q_{\psi} \dim_\psi \, Q_{\psi}+q_{\chi} \dim_\chi \, Q_{\chi} + \k_{u^{3}} Q_{u}$ & $-2835 p^{2} + \frac{8360}{3} p'^{2} - 2197500 s^{2}$ \\
  \hline
\end{tabular}
\caption{Anomalies of $SU(20), k=8$.}
\label{tab:anomalies-su 20 k 8}
\end{table}

Since $SU(2)_\psi$ is an anomaly-free group,  it does not need to break. The scenario that gives the lowest number of Goldstones amounts to breaking $SU(3)\chi\times U(1)_A$ to an anomaly-free subgroup. This can be achieved by condensing
\begin{eqnarray}
{\cal C}^{(ij)}=\psi^2\chi^{(i}\chi^{j)}\,,
\end{eqnarray}
which is singlet under $SU(2)_\psi$ and transforms in the $2$-index symmetric representation of $SU(3)$ breaking it to $SO(3)$. As before, this condensate also breaks $U(1)_A$ to the anomaly-free subgroup $\mathbb Z_2$. Thus, the IR unbroken $0$-form symmetry is $\frac{SU(2)\psi\times (\mathbb Z_2\subset U(1)_A)}{\mathbb Z_2}\times SO(3)$.

\section{Summary}
\label{Summary}
%
\begin{table}[h]
\begin{center}
\renewcommand*{\arraystretch}{1.75}
    \begin{tabular}{|l|l|l|l|}
    \hline
        Theory & Global Symmetries & Condensate(s) & IR Symmetries \\
        \hline
        $SU(5), k=1$ & $\frac{SU(9)_\chi\times U(1)_A}{\mathbb Z_5 \times \mathbb Z_9}$ &$\psi\chi^{9}\chi^{(i}\chi^{j)}$ & $SO(9) \times (\Z_{10}\subset U(1)_A)$\\
        $SU(6), k=1$ & $\f{SU(2)_{\psi} \times SU(10)_{\chi} \times U(1)_{A}}{\Z_{3} \times \Z_{2} \times \Z_{5}} \times \Z_{4}^{d\chi} $ & \textemdash & CFT \\
        $SU(6), k=2$ & $\frac{SU(5)_\chi\times U(1)_A}{\mathbb Z_3 \times \mathbb Z_5}\times \mathbb Z_4^{d\chi}$ & $ \psi^{2} \chi^{(i} \chi^{j)}$ & \makecell{$SO(5) \times$\\ $\frac{(\Z_{2}\subset U(1)_A) \times (\Z_{2}\subset \mathbb Z_4^{d\chi})}{\mathbb Z_2}$} \\
        $SU(10), k=2$ & $\f{SU(3)_{\psi} \times SU(7)_{\chi} \times U(1)_{A}}{\Z_{5} \times \Z_{3} \times \Z_{7}} \times \Z^{d\chi}_{4} $ & \textemdash & CFT \\\hline\hline
        
        $SU(8), k=2$ & $\frac{SU(2)_\psi\times SU(6)_\chi\times U(1)_A }{\mathbb Z_4\times \mathbb Z_6}\times \mathbb Z_{2}^{d\chi}$ & $\psi^2 \chi^{[i}\chi^{j]}$ & $\frac{SU(2)_\psi\times Sp(6)}{\mathbb Z_2}\times \mathbb Z_2^{d\chi}$ \\
        $SU(8), k=4$ & $\frac{SU(3)_\chi\times U(1)_A}{\mathbb Z_4\times \mathbb Z_{3}}\times \mathbb Z_2^{d\chi}$ & $ \psi\chi^i ,\psi^2\chi^{(i}\chi^{j)}$ &$SO(3)$ \\
        $SU(12), k=4$ & $\f{SU(2)_{\psi} \times SU(4)_{\chi} \times U(1)_{A}}{\Z_{6} \times \Z_{2} \times \Z_{2}}$ & $\psi_i\chi^j$ & $\frac{SU(2)_{V} \times (SU(2)\subset SU(4){\chi})}{\mathbb Z_2}$ \\
        $SU(12), k=8$ & $\f{SU(2)_{\chi} \times U(1)_{A}}{\Z_{3} \times \Z_{2}} $ & $\epsilon_{ij}\psi^2\chi^i\chi^j$ & $\frac{SU(2)_\chi \times (\Z_{2}\subset U(1)_A)}{\mathbb Z_2}$ \\
        $SU(16), k=4$ & $\f{SU(3)_{\psi} \times SU(5)_{\chi} \times U(1)_{A}}{\Z_{8} \times \Z_{3} \times \Z_{5}} \times \Z^{d\chi}_{2}$  & \textemdash  & CFT \\
        $SU(20), k=8$ & $\f{SU(2)_{\psi} \times SU(3)_{\chi} \times U(1)_{A}}{\Z_{5} \times \Z_{2} \times \Z_{3}} $ & $ \psi^2\chi^{(i}\chi^{j)}$ & $\frac{SU(2)_\psi \times (\Z_{2}\subset U(1)_A)}{\mathbb Z_2}\times SO(3)$ \\
        $SU(20), k=4$ & $\f{SU(4)_{\psi} \times SU(6)_{\chi} \times U(1)_{A}}{\Z_{10} \times \Z_{4} \times \Z_{3}} \times \Z^{d\chi}_{2}$  & \textemdash  & CFT \\
        $SU(28), k=8$ & $\f{SU(3)_{\psi} \times SU(4)_{\chi} \times U(1)_{A}}{\Z_{7} \times \Z_{3} \times \Z_{4}} $  & \textemdash  & CFT \\
        $SU(36), k=8$ & $\f{SU(4)_{\psi} \times SU(5)_{\chi} \times U(1)_{A}}{\Z_{18}}$  & \textemdash  & CFT \\
        $SU(44), k=8$ & $\f{SU(5)_{\psi} \times SU(6)_{\chi} \times U(1)_{A}}{\Z_{11} \times \Z_{5} \times \Z_{6}} $  & \textemdash  & CFT \\\hline
    \end{tabular}
\end{center}
\caption{A summary of the $2$-index chiral theories, their global symmetries, and their IR realizations. Theories with $N$ even also enjoy a $\mathbb Z_2^{(1)}$ $1$-form symmetry acting on the Wilson lines. This symmetry is assumed to be unbroken in theories that confine.}
\label{The IR realizations}
\end{table}

In this paper, we exhaustively scrutinized the $2$-index chiral gauge theories. By studying the $2$-loop and $3$-loop $\beta$-functions, we could pinpoint a few theories that may flow to an IR CFT. Theories that do not admit a fixed point break its global symmetries. We considered scenarios that give the minimal number of IR Goldstones, as this lowers the free energy of the theory. We paid particular attention to the anomaly-matching conditions and ensured that the condensates match any discrete subgroup of $U(1)_A$. Our theories, their global symmetries, the proposed IR phase condensates, and the unbroken IR symmetries are shown in Table \ref{The IR realizations}. The first $4$ theories are fermionic, while the rest are bosonic.

 Our investigation included a closer examination of the CFU anomalies one of the authors studied in the previous work \cite{Anber:2021iip}, giving a better interpretation of this class of anomalies in the light of the discrete-anomaly matching conditions. The general finding is that matching the full set of anomalies and, in particular, the anomalies of the discrete subgroups of the axial $U(1)_A$ symmetry necessitates the formation of multiple higher-order condensates. One expects such higher-order condensates to form in strongly-coupled theories. Here, their formation is explained via the constraints of the anomaly-matching conditions. We also employed a systematic approach to search for massless composite fermions that could match the anomalies in the case of fermionic theories. We were not able to find such composites. In one case, we used the CFU anomaly to show that a set of composites cannot solely match this anomaly, hinting at a deeper reason why the composites could not be found. 
 
Our work provides a systematic approach that can be applied to study other classes of strongly-coupled phenomena, including different chiral gauge theories.

{\bf {\flushleft{Acknowledgments:}}} We would like to thank Nakarin Lohitsiri and Erich Poppitz for various illuminating discussions and comments on the manuscript.  M.A. acknowledges the hospitality at the University of Toronto, where part of this work was completed.  This work is supported by STFC through grant ST/T000708/1.  

\section*{Appendix}
\appendix
\section{Obtaining the discrete chiral symmetry}
\label{Obtaining the Discrete Chiral Symmetry}

In this appendix, we show that there is a discrete symmetry $\mathbb Z_r$, where $r=\mbox{gcd}(N_\psi,N_\chi)$, that acts on $\chi$.
To this end, we conisder the groups $\Z_{N_{\psi}p_{\psi}+N_{\chi}p_{\chi}}$ and $U(1)_{A}$  we discussed in the text. 
Under $U(1)_{A} \times \Z_{N_{\psi}p_{\psi}+N_{\chi}p_{\chi}}$, $\psi$ transforms as
\begin{eqnarray}
\psi \longra e^{2\pi i aq_{\psi} \a} e^{2\pi i p_{\psi} \f{l}{N_{\psi}p_{\psi}+N_{\chi}p_{\chi}}} \psi\,,
\end{eqnarray}
where $l \in \Z_{N_{\psi}p_{\psi}+N_{\chi}p_{\chi}}$, $a$ is a charge factor, and $\a \in [0,1)$. This transformation leaves $\psi$ invariant if 
\begin{gather}
    aq_{\psi}\a + p_{\psi} \f{l}{N_{\psi}p_{\psi}+N_{\chi}p_{\chi}} = k_{1} \in \Z 
    \implies \a = \f{k_{1}}{aq_{\psi}} - \f{p_{\psi}l}{aq_{\psi} (N_{\psi}p_{\psi}+N_{\chi}p_{\chi})}\,.
\end{gather}
Note that $k_{1}$ can be freely chosen. Then, $\chi$ transforms under $U(1)_{A} \times \Z_{N_{\psi}p_{\psi}+N_{\chi}p_{\chi}}$ as:
\begin{eqnarray}
\nonumber
    \chi && \longra e^{2\pi i a q_{\chi}\a } e^{2\pi i p_{\chi}\f{l}{N_{\psi}p_{\psi}+N_{\chi}p_{\chi}}} \chi  = e^{2\pi i \left( aq_{\chi} \left( \f{k_{1}}{aq_{\psi}} - \f{p_{\psi}l}{aq_{\psi} (N_{\psi}p_{\psi}+N_{\chi}p_{\chi})}\right) + p_{\chi} \f{l}{N_{\psi}p_{\psi}+N_{\chi}p_{\chi}} \right)} \chi \\
    \nonumber
    && = e^{2\pi i \left( \f{q_{\chi}}{q_{\psi}} k_{1} + \f{l}{N_{\psi}p_{\psi}+N_{\chi}p_{\chi}} \left(p_{\chi} - p_{\psi} \f{q_{\chi}}{q_{\psi}} \right) \right)} \chi  = e^{2\pi i \left( \f{q_{\chi}}{q_{\psi}} k_{1} + \f{l}{(N_{\psi}p_{\psi}+N_{\chi}p_{\chi}) q_{\psi}} \left(-p_{\chi} \f{N_{\chi}}{r} - p_{\psi} \f{N_{\psi}}{r} \right) \right)} \chi\,,  \\
    && = e^{2\pi i \left( \f{q_{\chi}}{q_{\psi}} k_{1} - \f{l}{rq_{\psi}} \right)} \chi\,,
\end{eqnarray}
where $r = \gcd(N_{\psi}, N_{\chi})$ and we used $q_{\psi} = -\f{N_{\chi}}{r}$ and $q_{\chi} = \f{N_{\psi}}{r}$. We can rewrite $l = m_{1} + m_{2} r$, where $m_{1} = 0, 1, \dots, r-1$ and $m_2\in \mathbb Z$. Bezout's theorem also tells us that since $r = \gcd(N_{\psi}, N_{\chi})$, there are integers $k_{1}, k_{2}$ such that $m_{2}r = k_{1}N_{\psi} + k_{2} N_{\chi}$. Applying this to the transformation of $\chi$ gives us:
\begin{eqnarray}
\nonumber
    \chi && \longra e^{2\pi i \left( \f{q_{\chi}}{q_{\psi}} k_{1} - \f{l}{rq_{\psi}} \right)} \chi = e^{2\pi i \left( \f{q_{\chi}}{q_{\psi}} k_{1} - \f{m_{1} + m_{2} r}{rq_{\psi}} \right)} \chi 
    \nonumber
    = e^{2\pi i \left( \f{m_{1} + m_{2} r}{N_{\chi}} - \f{N_{\psi}}{N_{\chi}}k_{1} \right)} \chi  \\
    &&= e^{2\pi i \f{m_{1}}{N_{\chi}}} e^{2\pi i \f{m_{2}r - N_{\psi}k_{1}}{N_{\chi}}} \chi  = e^{2\pi i \f{m_{1}}{N_{\chi}}} e^{2\pi i k_{2}} \chi  = e^{2\pi i \f{m_{1}}{N_{\chi}}} \chi\,.
\end{eqnarray}
Since $m_{1} = 0, 1, \dots, r-1$, there are only $r$ distinct transformations generated by $U(1)_{A} \times \Z_{N_{\psi}p_{\psi}+N_{\chi}p_{\chi}}$, and the symmetry group that acts on $\chi$ is $\Z_{r}$. For our purposes, we will assume that under $\Z_{r}$, $\chi$ transforms with charge 1 (in principle, we could fix any charge). Finally, one needs to check whether this $\mathbb Z_r$ is a genuine symmetry in the sense that it cannot be absorbed in the center of color or flavor groups. This will be done on a case-by-case basis.

\section{The 3-loop $\beta$-function and the IR fixed points}
\label{2loop beta function}

The 3-loop $\beta$ function is given by (see \cite{Caswell:1974gg,Dietrich:2006cm,Zoller:2016sgq})
\begin{eqnarray}
\begin{split}
\beta(g)=&-\beta_0\frac{g^3}{(4\pi)^2}-\beta_1\frac{g^5}{(4\pi)^4}-\beta_2\frac{g^7}{(4\pi)^6}~,
\\[3pt]
\beta_0=&\frac{11}{6}C_2(G)-\sum_{\cal R}\frac{1}{3}T_{\cal R}n_{\cal R}~,
\\[3pt]
\beta_1=&\frac{34}{12}C_2^2(G)-\sum_{{\cal R}}\left\{\frac{5}{6}n_{\cal R} C_2(G)T_{\cal R} + \frac{n_{\cal R}}{2}C_2({\cal R})T_{\cal R}\right\}~,
\\[3pt]
\beta_2=&\frac{2857}{432}C_2^3(G)-\sum_{\cal{R}}\frac{n_{\cal R}T_{\cal R}}{4} \left[-\frac{C_2^2({\cal R})}{2}+\frac{205 C_2(G)C_2({\cal R})}{36}+\frac{1415C_2^2(G)}{108} \right]
\\[3pt]
&+\quad\sum_{\cal{R},\cal{R}'} \frac{n_{\cal R}n_{\cal R}' T_{\cal R}T_{\cal R'}}{16}\left[\frac{44 C_2({\cal R})}{18}+\frac{158C_2(G)}{54} \right]\,.
\end{split}
\label{beta function}
\end{eqnarray} 
Here, $G$ denotes the adjoint representation, and $n_{\cal R}$ is the number of the Weyl flavors in representation ${\cal R}$. Also, $C_2({\cal R})$  is the quadratic Casimir operator of representation ${\cal R}$,  defined as 
\begin{eqnarray}
t^a_{\cal R}t^a_{\cal R}=C_2({\cal R})\mathbf{1}_{\cal R}~.
\label{Casimir}
\end{eqnarray}
We reserve $C_2(G)$ for the quadratic Casimir of the adjoint representation. $T_{\cal R}$ is the Dynkin index of ${\cal R}$, which is defined by
\begin{eqnarray}
\mbox{tr}\left[t^a_{\cal R}t^b_{\cal R}\right]=T_{\cal R}\delta^{ab}~.
\label{trace}
\end{eqnarray}
From Eqs. (\ref{Casimir}) and (\ref{trace}), we easily obtain the useful relation 
\begin{eqnarray}
T_{\cal R}\mbox{dim}_G=C_2({\cal R})\mbox{dim}_{\cal R}~,
\end{eqnarray}
where $\mbox{dim}_{\cal R}$ is the dimension of $\cal R$.

In particular, we have $C_2(G)=2N$, $\mbox{dim}_G=N^2-1$, $T_\psi=N+2$, $\mbox{dim}_\psi=\frac{N(N+1)}{2}$, $C_2(\psi)=\frac{2(N+2)(N-1)}{N}$, $T_\chi=N-2$,  $\mbox{dim}_\chi=\frac{N(N-1)}{2}$, $C_2(\chi)=\frac{2(N-2)(N+1)}{N}$. Then, the values of $\beta_0$ to $\beta_2$ are 
\begin{eqnarray}
\begin{split}
\beta_0=&\ \frac{1}{3}\left[11N-\frac{2}{k}(N^2-8) \right]~,
\\[3pt]
\beta_1=&\ \frac{2\left(-48+76N^2+17kN^3-8N^4\right)}{3kN}~,
\\[3pt]
\beta_2=&\ \frac{1}{54k^2N^2}\Big [ 2857k^2N^5+N(-8448+12448N^2-2584N^4+145N^6)
\\[3pt]
&\quad -2k(864+3948N^2-8945N^4+988N^6)\Big ]~.
\end{split}
\end{eqnarray}
Assuming that $\beta_0>0$ and $\beta_1<0$, the theory develops an IR fixed point to $2$-loops. The value of the coupling constant at the fixed point is
\begin{eqnarray}
\alpha_{*}\equiv \frac{g_{*}^2}{4\pi}=-\frac{4\pi \beta_0}{\beta_1}=\frac{2\pi N \left(16+11kN-2N^2\right)}{48-76N^2-17kN^3+8N^4}~.
\end{eqnarray}
To assess the stability of this fixed point, we can examine the roots of the $\beta$-function when the 3-loop term is taken into account.

\bigskip

  \bibliography{References.bib}
  
  \bibliographystyle{JHEP}
  \end{document}